\documentclass[useAMS,usenatbib]{mn2e}
\usepackage{graphicx}
\usepackage{xcolor}
\usepackage{multirow}
\usepackage{wasysym}  
\usepackage{sidecap}
\usepackage{amssymb}
\title[UVES and X-Shooter spectroscopy of the emission line AM\,CVn systems GP Com and V396 Hya]{UVES and X-Shooter spectroscopy of the emission line AM\,CVn systems GP Com and V396 Hya}
\author[T. Kupfer et al.]{T. Kupfer$^{1,2}$\thanks{E-mail:tkupfer@caltech.edu}\thanks{Based on observations made with ESO telescopes at the Paranal Observatory under programme ID 69.D-0562(A) and 084.D-0814(A).}, D. Steeghs$^{3}$, P.~J. Groot$^{2}$, T.~R. Marsh$^{3}$, G. Nelemans$^{2,4}$ and \newauthor{G.~H.~A. Roelofs$^{2}$}\\
$^{1}$Division of Physics, Mathematics, and Astronomy, California Institute of Technology, Pasadena, CA 91125, USA\\
$^{2}$Department of Astrophysics/IMAPP, Radboud University Nijmegen, P.O. Box 9010, 6500 GL Nijmegen, The Netherlands\\
$^{3}$Department of Physics, University of Warwick, Coventry CV4 7AL, UK\\
$^{4}$Institute for Astronomy, KU Leuven, Celestijnenlaan 200D, 3001 Leuven, Belgium}

\begin{document}

\date{Accepted --- Received ---; in original form ---}

\pagerange{\pageref{firstpage}--\pageref{lastpage}} \pubyear{2002}

\maketitle

\label{firstpage}

\begin{abstract}
We present time-resolved spectroscopy of the AM\,CVn-type binaries GP\,Com and V396\,Hya obtained with VLT/X-Shooter and VLT/UVES. We fully resolve the narrow central components of the dominant helium lines and determine radial velocity semi-amplitudes of $K_{\rm spike} = 11.7\pm0.3$\,km\,s$^{-1}$ for GP\,Com and $K_{\rm spike} = 5.8\pm0.3$\,km\,s$^{-1}$ for V396\,Hya. The mean velocities of the narrow central components show variations from line to line. Compared to calculated line profiles that include Stark broadening we are able to explain the displacements, and the appearance of forbidden helium lines, by additional Stark broadening of emission in a helium plasma with an electron density $n_e\simeq 5\times 10^{15}$ cm$^{-3}$. More than $30$ nitrogen and more than $10$ neon lines emission lines were detected in both systems. Additionally, $20$ nitrogen absorption lines are only seen in GP Com. The radial velocity variations of these lines show the same phase and velocity amplitude as the central helium emission components.\\
The small semi-amplitude of the central helium emission component, the consistency of phase and amplitude with the absorption components in GP\,Com as well as the measured Stark effect shows that the central helium emission component, the so-called central-spike, is consistent with an origin on the accreting white dwarf.\\ 
We use the dynamics of the bright spot and the central spike to constrain the binary parameters for both systems and find a donor mass of $9.6$ - $42.8$\,M$_{\rm Jupiter}$ for GP\,Com and $6.1$ - $30.5$\,M$_{\rm Jupiter}$ for V396\,Hya.\\
We find an upper limit for the rotational velocity of the accretor of $v_{\rm rot}<46$\,km\,s$^{-1}$ for GP\,Com and $v_{\rm rot}<59$\,km\,s$^{-1}$ for V396\,Hya which excludes a fast rotating accretor in both systems. 

\end{abstract}

\begin{keywords}
accretion, accretion discs -- binaries: close -- stars: individual:  -- stars:
individual: GP Com, V396 Hya

\end{keywords}

\begin{table*}
 \centering
 \caption{Summary of the observations of V396\,Hya and GP\,Com}
  \begin{tabular}{clccccc}
  \hline
  System & Tele./Inst. & N$_{\rm exp}$ & Exp. time (s) &   Total time (h) & Coverage (\AA)   &  Resolution\\
  \hline\hline
  {\bf V396\,Hya} &     &         &  \\
  2002-04-07  & VLT/UVES (blue)  & 28 & 360  &    3.33 &    3826-5053  & $\sim$40\,000 \\
  2002-04-07  & VLT/UVES (red)  & 28 & 360  &   3.33 &     5763-9462  &  $\sim$40\,000 \\
  2010-02-18  & VLT/X-Shooter (UVB)  &  3 & 900 &   0.75  &    3050-5550 &   4300   \\ 
  2010-02-18  & VLT/X-Shooter (VIS) &  3 & 900 & 0.75  &    5340-10\,200 &   7400    \\ 
  2010-02-18  & VLT/X-Shooter (NIR  &  3 & 900 &  0.75  &   9940-24\,800 &   5400    \\ 
    \noalign{\smallskip}
  {\bf GP\,Com} &        &     &    \\
  2002-04-07  & VLT/UVES (blue)    & 110 & 120  &  5.25    &   3826-5053   &   $\sim$40\,000 \\
    2002-04-07  & VLT/UVES (red)      & 110 & 120  &  5.25    &   5763-9462   &  $\sim$40\,000 \\
  2010-02-18  & VLT/X-Shooter (UVB)  & 45  &  60  &  1.00    &   3050-5550   &   4300  \\
  2010-02-18  & VLT/X-Shooter (VIS) & 45  &  60  & 1.00    &  5340-10\,200   &   7400  \\
  2010-02-18  & VLT/X-Shooter (NIR) & 45  &  60  &  1.00    &   9940-24\,800  &   5400  \\

   \hline
\end{tabular}
\label{observ}
\end{table*}

\section{Introduction}

Accretion onto white dwarfs in cataclysmic variables (CVs) commonly proceeds via Roche-lobe overflow of hydrogen-rich material from a main-sequence type donor star. A small number of systems have been identified with hydrogen-deficient, degenerate, donor stars; the AM\,CVn systems (see \citealp{sol10} for a recent review). These systems consist of a white dwarf (WD) primary and either a WD or significantly evolved semi-degenerate helium star companion (e.g. Nelemans et al. 2001)\nocite{nel01}. Observationally they are characterized by a high deficiency of hydrogen ($<1\,\%$) and short orbital periods ($<$\,1\,hour), indicating an advanced stage of binary evolution.  

Since the identification of the prototype, AM\,CVn, as a semi-detached pair of degenerate dwarfs \citep{sma67,pac67,fau72}, over 40 additional systems and candidates have been identified. Confirmed orbital periods range from 5-65 minutes (e.g \citealp{nat81,odo87,odo94,rui01,wou03,roe05,roe06,roe07a,and05,and08,roe10,lev11,kup13,lev13,car14}). AM\,CVn binaries are important as strong, low-frequency, Galactic gravitational wave sources (e.g. \citealp{nel04, roe07b, nis12}), the source population of the proposed €.Ia supernovae \citep{bil07}, and as probes of the final stages of low-mass binary evolution.



Several formation channels have been proposed for these systems; a double white dwarf channel \citep{tut79}, a channel in which the donors are low-mass helium stars \citep{sav86,tut89,yun08}, and one with evolved post-main-sequence donors \citep{tho02,pod03}.  A way to distinguish between these scenarios is to obtain the chemical composition of the donor, and in particular the C/O, N/O and N/C ratios, due to different levels of CNO and He burning in the progenitor of the donor \citep{nel10}. High N/O and N/C ratios are expected for a helium white dwarf donor and significantly lower N/O and N/C ratios expected for a semi-degenerate donor. Although the donor has never been observed directly, the accreted material in the disc and the photosphere of the accreting white dwarf is expected to represent the composition of the donor. 

In this paper we present high-resolution optical spectroscopy of the AM\,CVn systems GP Com \citep{nat81} and V396 Hya \citep{rui01}. These two systems have relatively long orbital periods (46 and 65 minutes) and represent, evolutionary speaking, the bulk of the AM\,CVn systems, which should be long period, low mass-transfer-rate objects ($\sim 10^{-12} M_{\odot}$\,yr$^{-1}$).  In these systems, the mass-transfer-rate was thought to be below the threshold for accretion disc instabilities to occur \citep{hir90}, and indeed no outburst has so far been reported for GP Com or V396 Hya. 

GP Com has been observed extensively, and shows erratic flaring in optical, UV \citep{mar95} and X-ray \citep{van94} wavebands, which is attributed to accretion. One of the more intriguing spectral features is the presence of sharp, low-velocity components in the optical helium emission lines. The fact that these `central-spikes' contribute to the flare spectrum and follow a low-amplitude radial velocity curve as a function of the orbital phase suggests an origin on, or near, the white dwarf accretor \citep{mar99,mor03}. Such so-called `central-spikes' have so far only been observed in AM\,CVn systems and He-rich dwarf novae but never in hydrogen dominated cataclysmic variables \citep{bre12}. If the accretor origin of the central-spikes can be confirmed, the central-spikes would be a powerful tool to trace the motion of the accreting white dwarf. 

The accreting white dwarf accretes helium with a specific angular momentum from the inner edge of the accretion disc. This will spin up the accretor and is expected to result in a minimum equatorial velocity of $\sim1250$\,km\,s$^{-1}$ \citep{bil06}. The rapid surface rotation will broaden any spectral features originating on the white dwarf significantly. So far no detailed study of the rotational velocity has been performed. However, the narrow central-spike features already suggest a much lower rotational velocity, if the origin of the central-spike on the white dwarf can be confirmed. This would imply an effective loss of angular momentum from the rapidly spinning accretor.

In an attempt to understand the origin of the peculiar spike, we secured echelle spectroscopy of GP\,Com and V396\,Hya. The UVES and X-Shooter spectra allow us to fully resolve the kinematics of the line profiles in general, and their central-spikes in particular.

\begin{figure*}
\begin{center}
\includegraphics[width=0.99\textwidth]{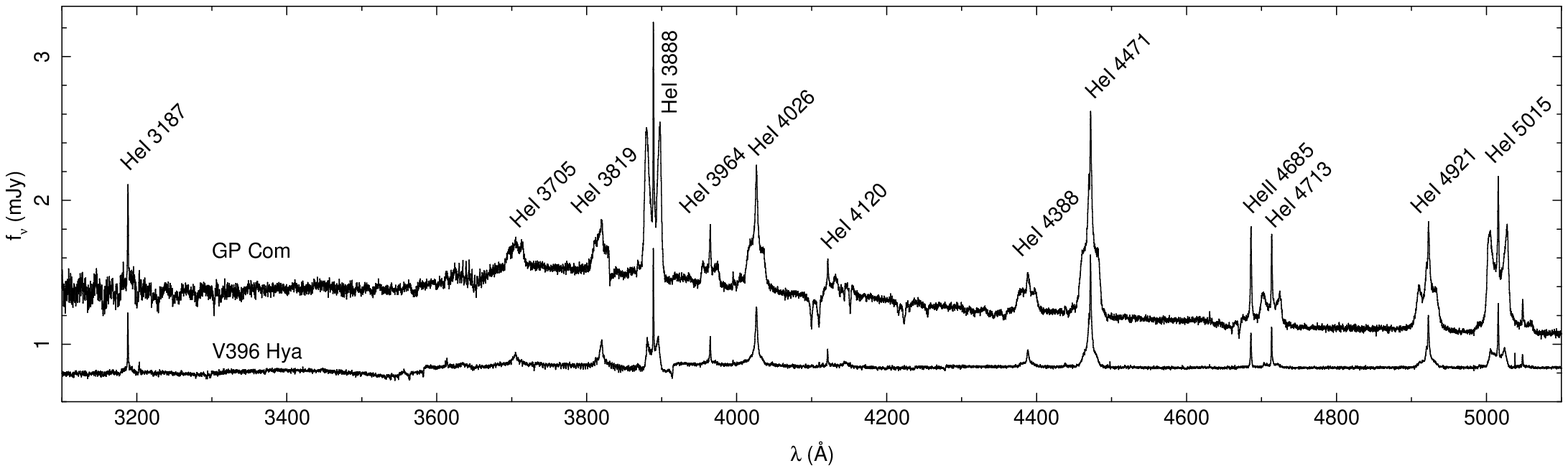}
\includegraphics[width=0.99\textwidth]{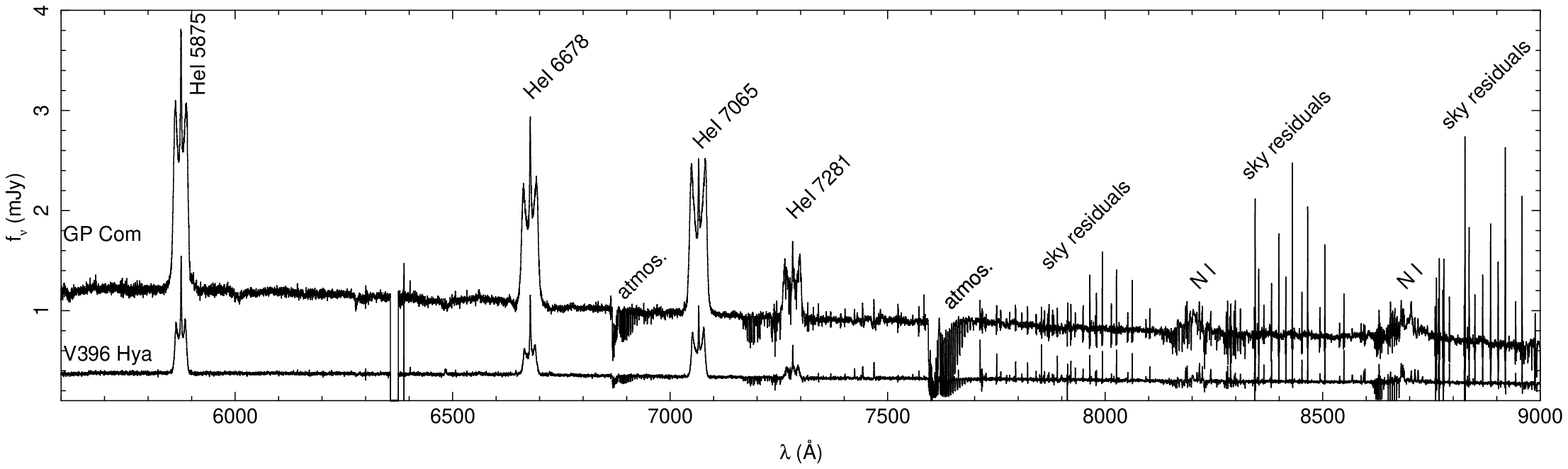}
\includegraphics[width=0.47\textwidth]{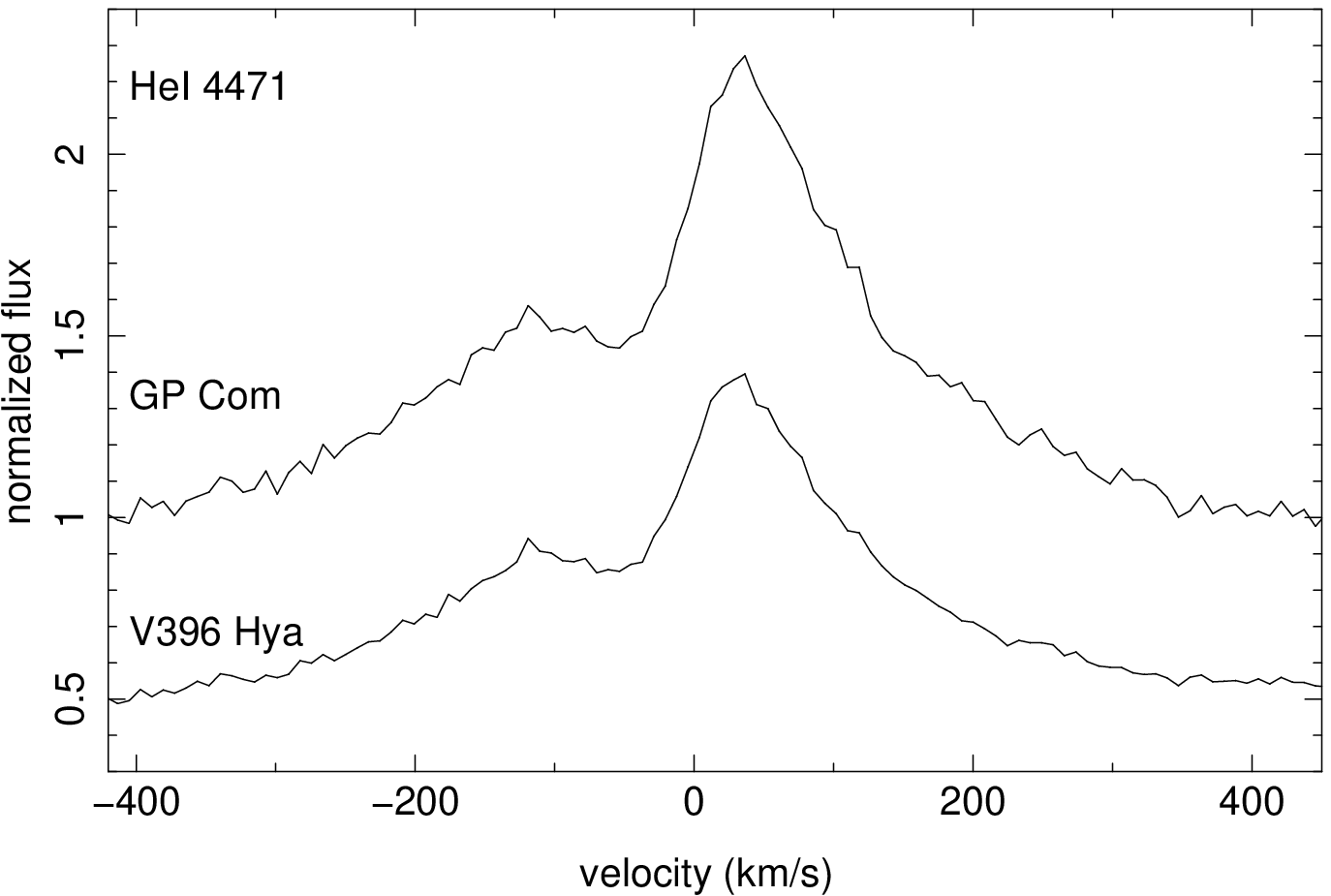}
\includegraphics[width=0.48\textwidth]{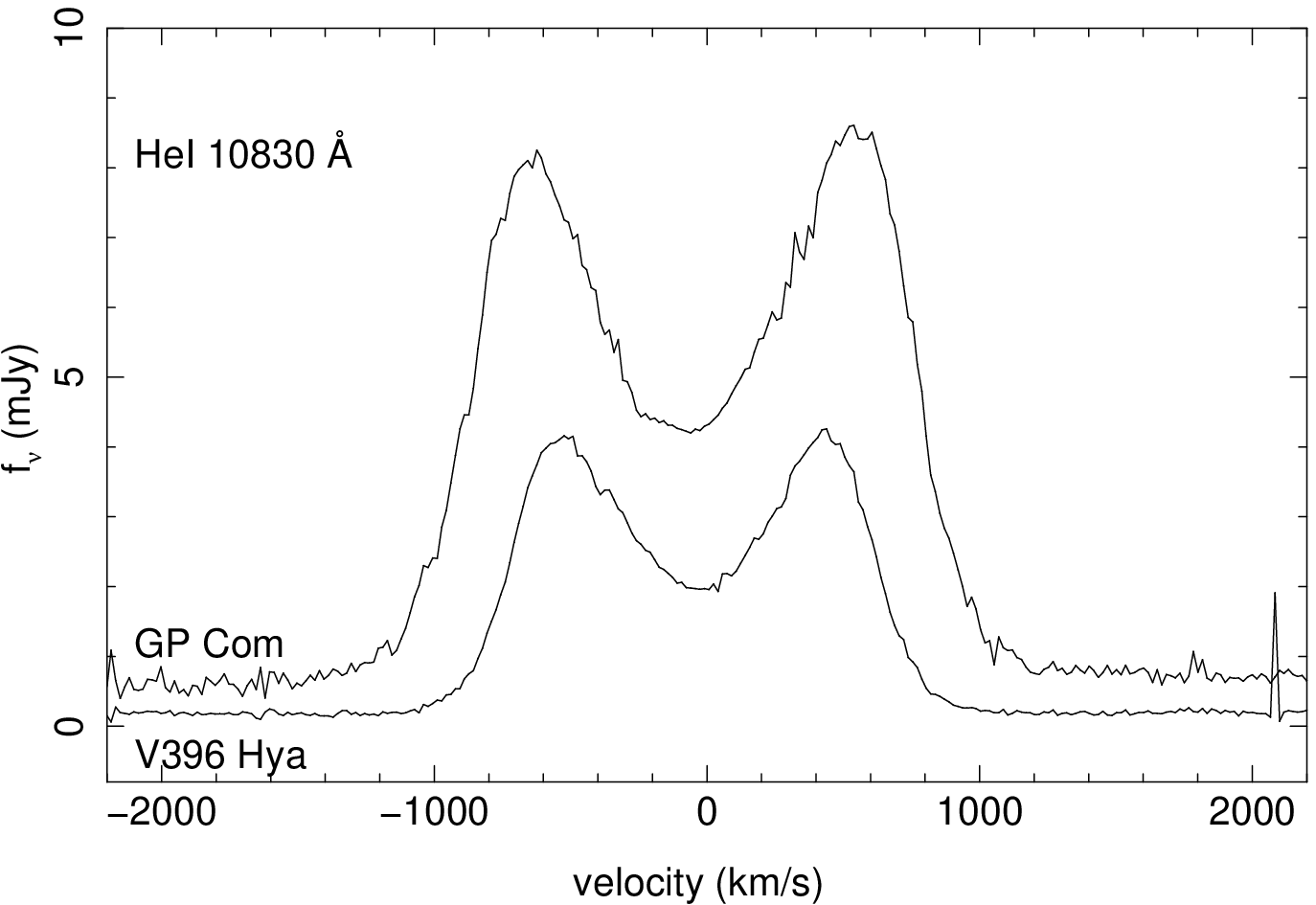}
\caption{Gaussian smoothed average spectrum of GP\,Com and V396\,Hya obtained with VLT/X-Shooter. Helium emission lines are indicated. The lower left panel shows the central-spike feature with its blue-shifted forbidden component in He\,{\sc i} 4471\,\AA\, observed with UVES. The lower right panel shows the strongest helium lines in the NIR arm observed with X-Shooter. The narrow spikes at wavelength $>\,7500$\,\AA\, are residuals from the nights sky line removal.}
\label{xshootaver}
\end{center}  
\end{figure*}


\section{Observations and Data reduction}

\subsection{VLT/UVES observations}


\begin{figure*}
\begin{center}
\includegraphics[width=0.24\textwidth]{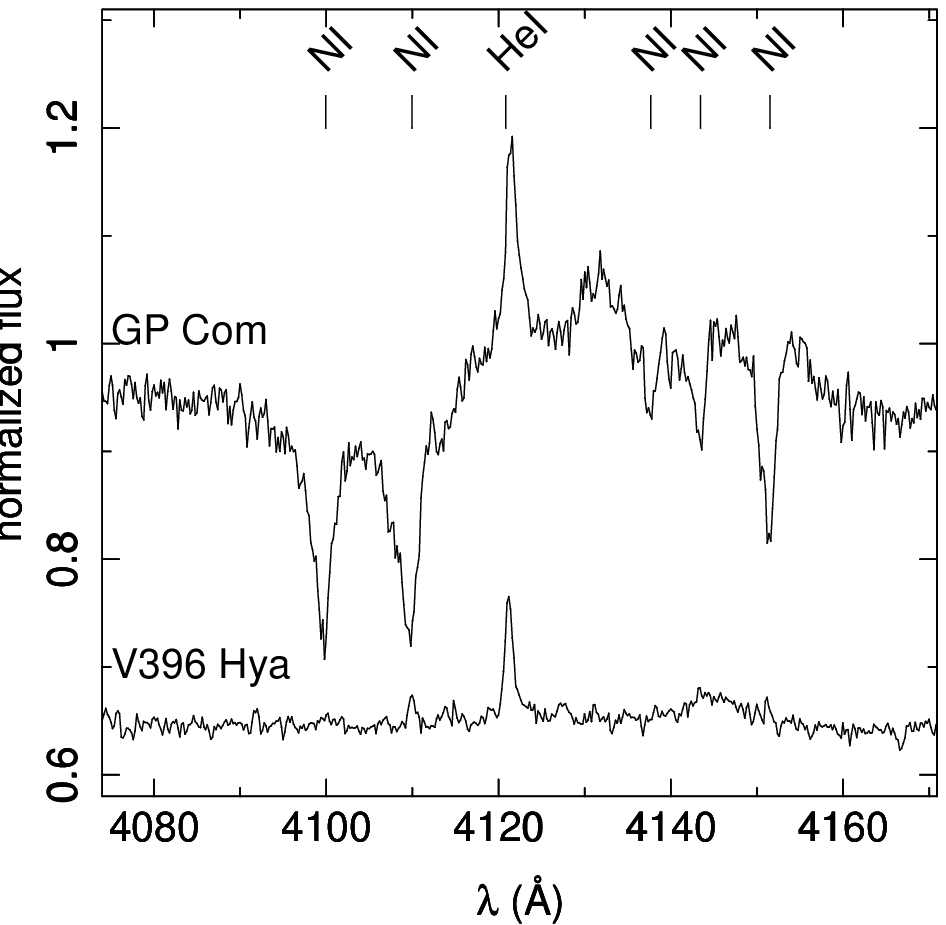}
\hspace*{0.01cm}
\includegraphics[width=0.24\textwidth]{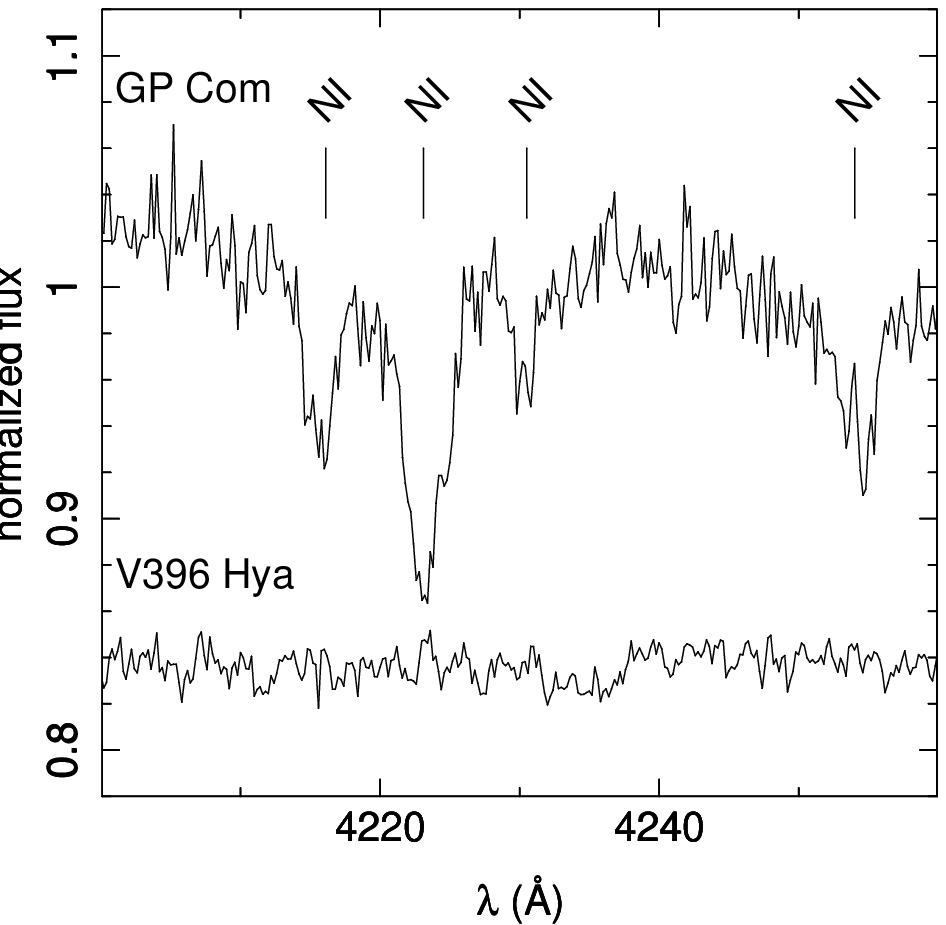}
\hspace*{0.01cm}
\includegraphics[width=0.24\textwidth]{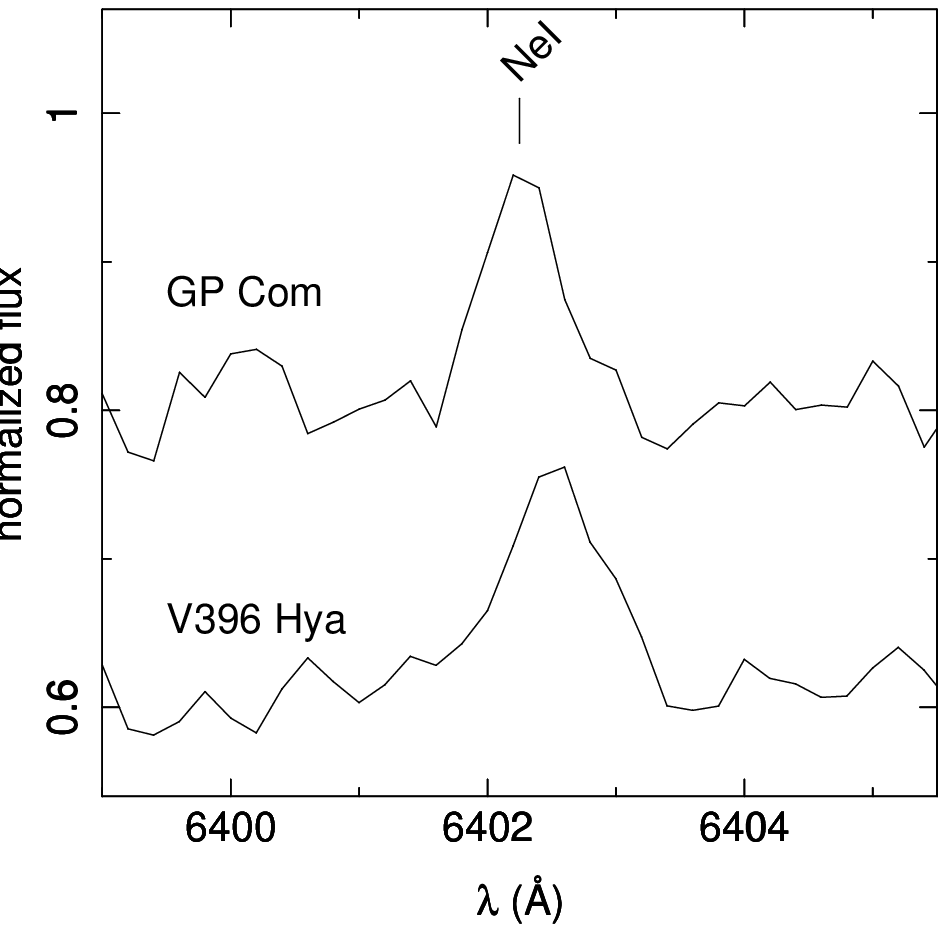}
\hspace*{0.01cm}
\includegraphics[width=0.24\textwidth]{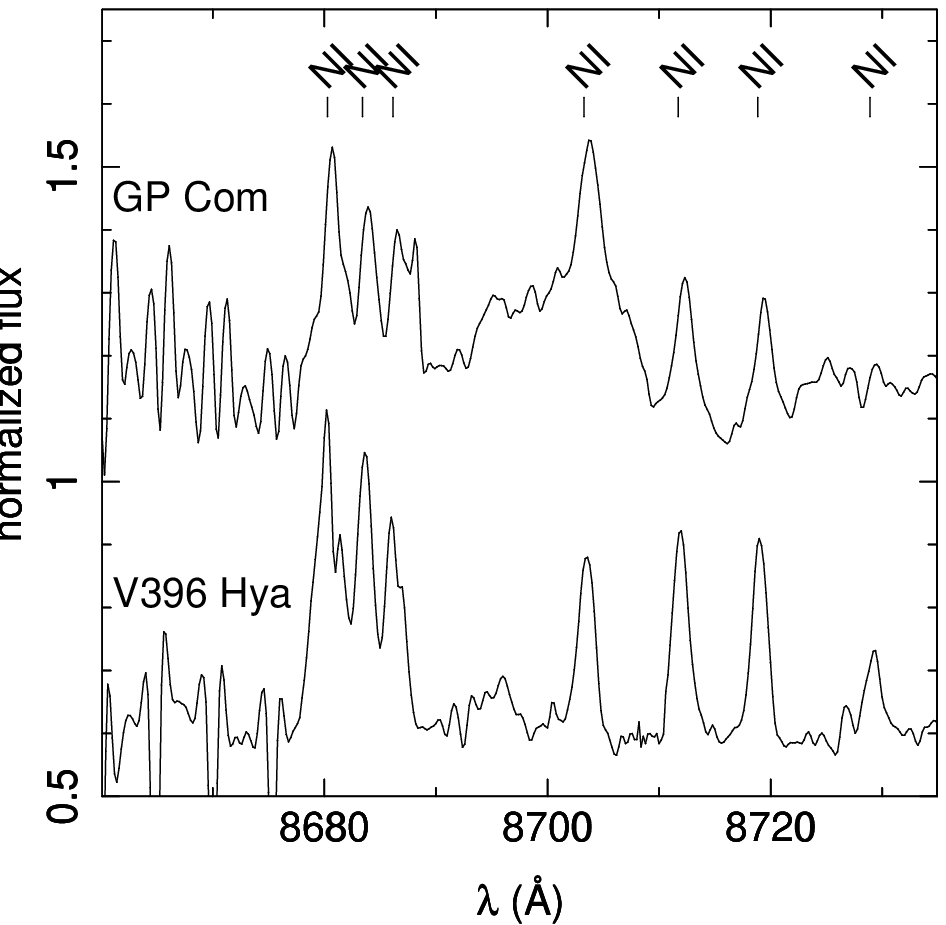}
\caption{Zoomed region of the average X-Shooter spectra around locations of various metal lines in GP\,Com and V396\,Hya.}
\label{metal}
\end{center}  
\end{figure*}

\begin{table}
\begin{center}
\caption{Emission line properties for GP Com}
\begin{tabular}{lrrrr}
\hline
 & \multicolumn{2}{c}{UVES} & \multicolumn{2}{c}{X-Shooter} \\
He line &  EW range  & EW mean & EW range & EW mean \\
         &     (\AA)  & (\AA)  & (\AA)  & (\AA)   \\
\hline\hline
3187.745 &    -     &    -    &  0.5-1.6 & 1.0 \\  
3705.005 &    -     &    -    &  2.6-3.8 & 3.2 \\  
3819.607 &    -     &    -    &  1.7-4.0 & 3.0 \\
3871.791 & \multirow{2}{*}{6.2-11.1$^{a}$} & \multirow{2}{*}{8.6$^{a}$}  &   \multirow{2}{*}{10.9-14.9$^{a}$} &    \multirow{2}{*}{12.6$^{a}$}\\
3888.643 &           &    &   &  \\
3964.730 & 0.7-2.1  &    1.4 & 1.2-2.3  & 1.8 \\
4026.191 & 4.6-8.7  &    6.6 & 5.2-9.5  & 7.3 \\
4387.929 & 1.1-2.9  &    1.9 & 2.9-4.9  & 3.9 \\
4471.502 & 8.3-14.9 &    11.0 & 11.7-20.3 & 14.8 \\
4685.710 & \multirow{2}{*}{3.5-6.5$^{b}$}  &   \multirow{2}{*}{4.9$^{b}$}  & \multirow{2}{*}{5.9-9.0$^{b}$}  &  \multirow{2}{*}{7.2$^{b}$} \\
4713.170 &   &  &&\\
4921.930 & 5.0-8.7 &    6.8 & 8.1-12.4  & 10.4 \\
5015.678 & \multirow{2}{*}{9.9-14.8$^{c}$} &    \multirow{2}{*}{12.4$^{c}$} &  \multirow{2}{*}{17.4-22.2$^{c}$}  &   \multirow{2}{*}{20.1$^{c}$} \\
5047.738 &  &  &&\\
5875.661 & 30.4-62.7 & 44.1 &  46.7-58.8  & 52.4 \\
6678.152 & 24.8-38.1 & 30.7 &  36.9-46.1  & 42.5 \\
7065.251 & 35.3-51.0 & 41.5 &  47.8-56.2  & 51.5 \\
7281.351 & 11.0-36.5 & 18.2 &  16.1-19.4  & 18.2 \\
10\,830.34 &   -       &  -   &  581.0-736.6 & 666.9 \\
12\,784.79 &   -       &  -   &     $^d$     &  $^d$  \\
17\,002.47 &   -       &  -   &     $^d$     &  $^d$  \\
20\,586.92 &   -       &  -   &     $^d$     &  $^d$  \\
\hline 
\label{tab:equiwidth_gpcom}
\end{tabular}
\begin{flushleft}
$^a$ Combined equivalent width of He\,{\sc i} 3871 and He\,{\sc i} 3888\\
$^b$ Combined equivalent width of He\,{\sc ii} 4685 and He\,{\sc i} 4713\\
$^c$ Combined equivalent width of He\,{\sc i} 5015 and He\,{\sc i} 5047\\
$^d$ Line present but contaminated with atmosphere.
\end{flushleft}
\end{center}
\end{table}

\begin{table}
\begin{center}
\caption{Emission line properties for V396 Hya}
\begin{tabular}{lrrrr}
\hline
 & \multicolumn{2}{c}{UVES} &\multicolumn{2}{c}{X-Shooter} \\
He line &  EW range  & EW mean  & EW range  & EW mean \\
         &     (\AA)  & (\AA)  &  (\AA)   & (\AA) \\
\hline\hline
3187.745 &    -     &    -    &   3.4-4.0 &  3.8  \\  
3705.005 &    -     &    -    &   2.0-2.5  & 2.2   \\  
3819.607 &    -     &    -    &   5.2-5.7  & 5.5  \\
3871.791 & \multirow{2}{*}{14.0-27.9$^{a}$} & \multirow{2}{*}{20.8$^{a}$} &  \multirow{2}{*}{15.9$^{a}$} &\multirow{2}{*}{15.9$^{a}$} \\ 
3888.643 &   &    &   &  \\
3964.730 &  2.8-5.2  & 3.7  &   1.9-2.6  &  2.2 \\
4026.191 &  16.4-25.4   & 20.9  &  10.3-11.4  &  11.0 \\
4387.929 &  4.6-7.7 & 6.1  &  3.0-3.5 & 3.3 \\
4471.502 &  24.4-39.0 & 31.1  &  17.5-19.6  & 18.2 \\
4685.710 &  \multirow{2}{*}{8.1-14.4$^{b}$} & \multirow{2}{*}{11.1$^{b}$}  &  \multirow{2}{*}{5.4-6.5$^{b}$} & \multirow{2}{*}{5.8$^{b}$} \\
4713.170 &   &  & \\
4921.930 &  11.6-20.1 & 15.3  &  8.6-9.8  & 9.1 \\
5015.678 &   \multirow{2}{*}{17.5-29.9$^{c}$} &  \multirow{2}{*}{22.7$^{c}$} &   \multirow{2}{*}{13.2-13.9$^{c}$} & \multirow{2}{*}{13.6$^{c}$}\\
5047.738 &  &  &\\
5875.661 &  58.3-101.8 & 76.4  &  50.4-53.7 & 51.5 \\
6678.152 &  45.0-71.7 & 57.2  &    32.8-35.9  & 34.6 \\
7065.251 &  54.2-89.1 & 70.9   &   48.6-51.7  &  50.6 \\
7281.351 &  19.3-38.0 & 24.5   &  12.6-13.7  & 13.1 \\
10\,830.34 &   -       &  -    &  922.3-982.6 & 969.5 \\
12\,784.79 &   -       &  -   &   $^d$   &  $^d$    \\
17\,002.47 &   -       &  -   &   $^d$  &  $^d$   \\
20\,586.92 &   -       &  -   &   $^d$   &   $^d$    \\
\hline
\label{tab:equiwidth_ce315}
\end{tabular}
\begin{flushleft}
$^a$ Combined equivalent width of He\,{\sc i} 3871 and He\,{\sc i} 3888\\
$^b$ Combined equivalent width of He\,{\sc ii} 4685 and He\,{\sc i} 4713\\
$^c$ Combined equivalent width of He\,{\sc i} 5015 and He\,{\sc i} 5047\\
$^d$ Line present but contaminated with atmosphere.
\end{flushleft}
\end{center}
\end{table}

We employed the UV-Visual Echelle Spectrograph (UVES) mounted on the Unit 2 telescope of the ESO-VLT at Cerro Paranal, Chile. The data were obtained in visitor mode during the night of April 7/8, 2002. The instrument was configured in dichroic mode (dichroic \#2) permitting simultaneous data acquisition from both the red and the blue camera. No pre-slit optics were put in the beam, instead atmospheric dispersion was minimized by maintaining the 1$''$ wide slit at the parallactic angle throughout the night. On the blue camera, a 2048x4096 pixel EEV detector covered 3826--5053\AA~in 29 echelle orders. The chip was read out using 1 port and 2x3 on-chip binning to reduce read-out time to 22s. In this mode, a binned pixel corresponds to 4.5 km/s along the dispersion axis and a slit-limited resolution profile of 1.7 pixels. The red camera detector system consists of two 2048x4096 pixel CCD detectors, an EEV device covering 5763--7502\AA~in 24 orders and a red optimized MIT-LL chip covering 7760--9462\AA~in 15 orders. Both chips were readout in 32s using 1 port each with 2x3 on-chip binning. The spectra pixel scale corresponds to 3.6 km/s after binning.

Frames were first de-biased using a median of 9 bias exposures. In addition, any residual bias contribution was subtracted using the overscan areas of the three CCD detectors. The orders were extracted using the \textsc{echomop} echelle package \citep{mil14}. Flat field correction was performed using the median of a series of well-exposed tungsten flat field exposures obtained during the day. The individual orders were traced and profile weights were determined for optimal extraction \citep{hor86}. Target exposures were then optimally extracted after subtracting the sky background. In addition, the same profile weights were used to extract a suitable ThAr exposure as well as the median flat-field frame. The wavelength scales were determined by fitting 4-6 order polynomials to reference lines in the extracted ThAr spectrum, delivering zero-point RMS residuals of $\le$0.6\,km/s. The object orders were blaze corrected by division of a smoothed version of the extracted flat field spectrum. Orders were then merged into a single spectrum using inverse variance weights based on photon statistics to combine overlapping order segments. Finally, the merged spectra were flux calibrated using wide slit exposures of the B-star flux standard HD\,60753.

For the red arm data, we also performed a correction for the telluric absorption features. Spectra were first aligned by cross-correlating with the telluric standard. This corrected small wavelength shifts and corrections were found to be less than 3 km/s. The telluric features in each spectrum were then removed as far as possible by adjusting the depth of the telluric template.

\begin{figure*}
\centering
\includegraphics[width=\textwidth]{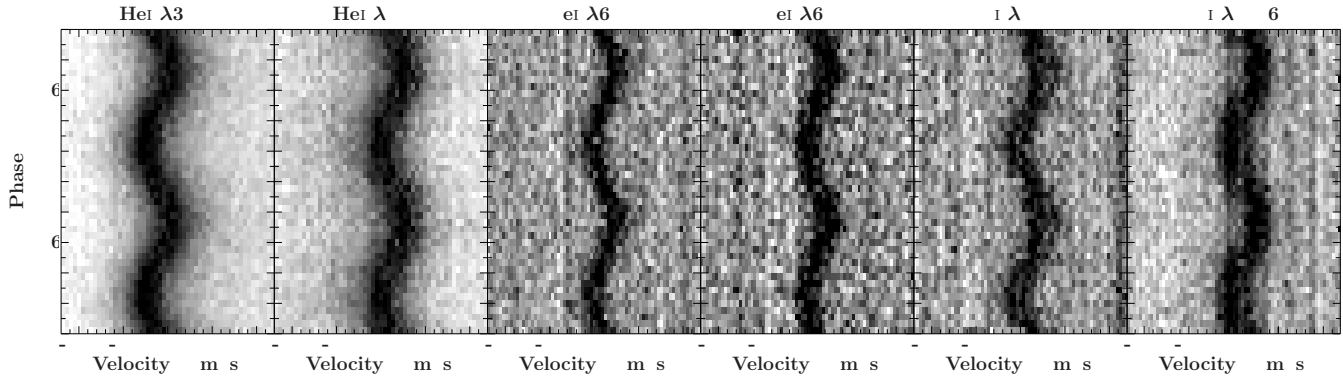}
\caption{Tracing of the central-spike of selected helium lines and some neon and nitrogen emission lines of GP Com from UVES data. \label{fig:trail_emis_gpcom}}
\end{figure*}

\begin{table}
\begin{center}
\caption{Velocity of the central-spike}
\begin{tabular}{lrrrr}
\hline
 & GP Com && V396 Hya& \\
HeI line & $\gamma$ & $K_1$ & $\gamma$ & $K_1$ \\
  (\AA)       &  (km\,s$^{-1}$) &  (km\,s$^{-1}$) & (km\,s$^{-1}$) & (km\,s$^{-1}$) \\
\hline \hline
3888.643 & --4.7 $\pm$ 0.2 & 12.7 $\pm$ 0.3 & --9.9 $\pm$ 0.3 & 6.1 $\pm$ 0.4\\
3964.730 & --4.1 $\pm$ 1.1 & 13.0 $\pm$ 1.6 & --2.5 $\pm$ 1.2 & 5.8 $\pm$ 1.6 \\
4387.929 & 14.5 $\pm$ 2.3 & 11.5 $\pm$ 3.2 & 20.1 $\pm$ 3.4 & 8.6 $\pm$ 4.2 \\
4471.502 & 42.4 $\pm$ 0.5 & 11.1 $\pm$ 0.7 & 39.4 $\pm$ 0.6 & 5.4 $\pm$ 0.9\\
4685.710 & 17.4 $\pm$ 0.3 & 11.7 $\pm$ 0.5 & 16.1 $\pm$ 0.5 & 5.2 $\pm$ 0.6\\
4713.170 & 32.6 $\pm$ 0.3 & 11.1 $\pm$ 0.5 & 27.3 $\pm$ 0.5 & 4.3 $\pm$ 0.6\\
4921.930 & 52.6 $\pm$ 1.0 & 12.3 $\pm$ 1.6 & 47.8 $\pm$ 0.8 & 5.8
$\pm$ 1.1\\
5015.678 & 6.4  $\pm$ 0.3 & 12.3 $\pm$ 0.4 & 7.5 $\pm$ 0.4 & 6.0 $\pm$
0.5\\
5875.661 & 0.4  $\pm$ 0.3 & 11.3 $\pm$ 0.3 & 2.5 $\pm$ 0.3 & 5.1 $\pm$
0.5\\
6678.152 & 18.2 $\pm$ 0.2 & 11.6 $\pm$ 0.2 & 11.2 $\pm$ 0.3 & 6.8
$\pm$ 0.3\\
7065.251 & 23.2 $\pm$ 0.2 & 13.3 $\pm$ 0.3 & 16.0 $\pm$ 0.3 & 6.2 $\pm$ 0.4 \\
7281.351 & 17.2 $\pm$ 0.2 &  8.0 $\pm$ 0.4 & 16.3 $\pm$ 0.4 & 4.1
$\pm$ 0.5 \\
\hline
Mean & & 11.7 $\pm$ 0.3 & & 5.8 $\pm$ 0.5 \\
\hline
\end{tabular}
\label{spikevelocities}
\end{center}
\end{table}
\subsection{VLT/X-Shooter observations}
GP\,Com and V396\,Hya were also observed using the medium resolution spectrograph X-Shooter \citep{ver11} mounted at the Cassegrain focus on the Unit 2 telescope of the VLT array in Paranal, Chile, during the night of 2010-02-18 as part of the Dutch GTO program. X-Shooter consists of 3 independent arms that give simultaneous spectra longward of the atmospheric cutoff (0.3 microns) in the UV (the `UVB' arm), optical (the `VIS' arm) and up to 2.5 microns in the near-infrared (the `NIR' arm). We used slit widths of 1.0$''$, 0.9$''$ and 0.9$''$ in X-Shooter's three arms and binned by 2x2 in the UVB and VIS arms resulting in velocity resolutions of 14\,km\,s$^{-1}$,  7\,km\,s$^{-1}$ and 11\,km\,s$^{-1}$ per binned pixel for the UVB, VIS and NIR arm respectively. The reduction of the raw frames was conducted using the standard pipeline release of Reflex (version 2.3) for X-Shooter data \citep{fre13}. The standard recipes were used to optimally extract and wavelength calibrate each spectrum. The instrumental response was removed by observing the spectrophotometric standard star EG\,274 \citep{ham92,ham94} and dividing it by a flux table of the same star to produce the response function.

Table\,\ref{observ} gives an overview of all observations and the instrumental set-ups.

\section{Analysis}

\subsection{Average spectra}
We present the time averaged spectra of GP Com and V396 Hya in Fig.\,\ref{xshootaver}, illustrating the striking similarity between the spectra of these two systems. The familiar triple-peaked helium emission lines dominate the spectra. The high-resolution spectra obtained with UVES nicely resolve the central-spike feature in both systems (lower left panel Fig.\,\ref{xshootaver}). We measured the equivalent widths (EW) for a number of prominent lines in all spectra. In Table \ref{tab:equiwidth_gpcom} and \ref{tab:equiwidth_ce315} we list the ranges of EWs observed together with their mean value. Large intrinsic EW modulations are detected in all lines throughout our observing run. In the case of GP Com, the measured EWs from UVES and X-Shooter are consistently smaller than previously reported values. In comparison with \citet{mar91}, we find that our mearured EWs are smaller by a factor between $1.4$ for He\,{\sc i}\,$7281$\AA\  and $3.3$ for He\,{\sc i}\,$4387$\AA\ respectively. Thus, on top of short timescale flaring activity, GP Com also displays significant variability on longer timescales. V396\,Hya shows very similar variations, with EW variations close to a factor of two for UVES.  We calculate flare spectra according to the methods described in \citet{mar95} and find that in both GP Com and V396 Hya the continuum as well as the disc and spike components of the lines contribute to the flaring. As discussed in \citet{mar99} and \citet{mor03}, this strongly suggests that the spike originates at or near the accreting dwarf and not in an extended nebula in both of these objects.

Despite the detailed similarities between the profile shapes in GP Com and V396 Hya, a few differences can be identified in the average spectra. GP Com displays a number of sharp absorption features, most notably between 4050\,\AA\,and 4250\,\AA. We identify these lines as N\,{\sc i}. No such features are present in V396 Hya (Fig.\,\ref{metal}). A large number of N\,{\sc ii} absorption lines have been observed in the high state system SDSS\,J1908 \citep{kup15}. \citet{mor03} already reported an absorption feature near He\,{\sc ii} 4686\,\AA\,in GP\,Com. Indeed this feature is also present in our X-Shooter and UVES spectra but its origin remains unidentified. 


In the blue spectral region the strongest N\,{\sc ii} lines are detected in absorption whereas in the red spectral region, complex N\,{\sc i} emission dominates the spectrum. A large number of individual transitions of N\,{\sc i} can be identified thanks to our high spectral resolution (Fig.\,\ref{metal}). The equivalent widths of these blends of N\,{\sc i} are larger in V396\,Hya than in GP\,Com. We also identify a number of weak Ne\,{\sc i}  emission features in the spectra of both systems (see Fig.\,\ref{metal} and \ref{fig:trail_emis_gpcom}). See Tab.\,\ref{tab:equi_1} and \ref{tab:equi_2} for an overview of the detected lines with measured equivalent widths.

The X-Shooter spectra allow us to search for spectroscopic signatures in the near-infrared part of the spectrum. We find a strong He\,{\sc i}\,$10\,830$\AA\, line with an equivalent width of $666.9$\,\AA\, and $969.5$\,\AA, which is more than ten times stronger than the strongest optical helium lines. Additionally we find the He\,{\sc i} lines $11\,969, 12\,784, 17\,002$ and $20 \,586$\,\AA. None of the helium lines in the near-infrared spectrum of GP\,Com and V396\,Hya show a sharp central-spike feature (lower right panel Fig.\,\ref{xshootaver}), a somewhat surprising difference to the optical regime. A possible explanation could be that the double-peaked helium disc lines in the near-infrared form at a lower temperature compared to the central-spike. In this case the near-infrared helium disc lines become stronger relative to the central-spike features and possibly outshine the central-spikes in the near-infrared.

The peak-to-peak velocity of the double-peaked disc emission line He\,{\sc i}\,10\,830\AA\, was found to be $1237\pm7$\,km\,s$^{-1}$ for GP\,Com and $1069\pm6$\,km\,s$^{-1}$ for V493\,Hya which is significantly lower than the $1379$-$1414$\,km\,s$^{-1}$ range for GP\,Com and the $1111$-$1124$\,km\,s$^{-1}$ range for V493\,Hya using the lines in the optical regime, showing an emission profile that is more centered in the outer disc.



\begin{figure}
\begin{center}
\includegraphics[width=0.48\textwidth]{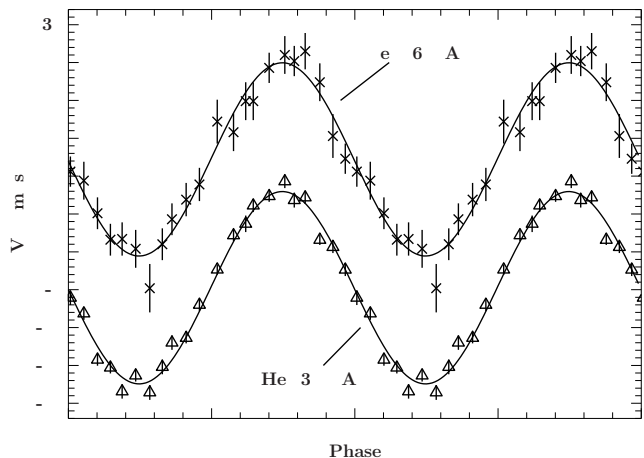}
\caption{Radial velocity curves of the central-spike of He\,{\sc i} 3888\,\AA\, and the emission line Ne\,{\sc i} 6402\,\AA\, in GP\,Com. Two orbits are plotted for better visualization.}
\label{fig:vel_emission_gpcom}
\end{center}
\end{figure}

\begin{figure}
\begin{center}
\includegraphics[width=0.48\textwidth]{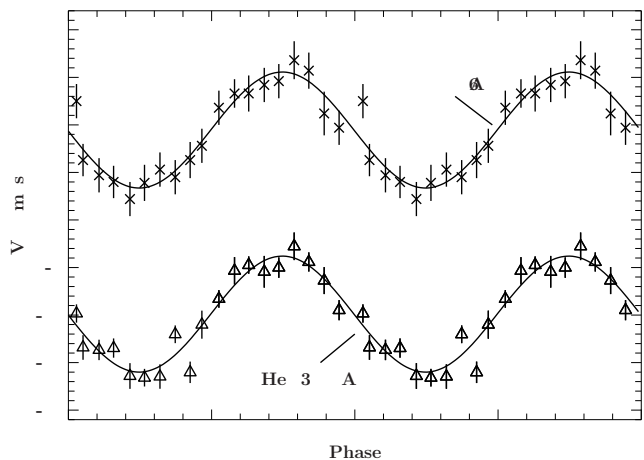}
\caption{Radial velocity curves of the central-spike of He\,{\sc i} 3888\,\AA\, and the emission line N\,{\sc i} 7468\,\AA\, in V396\,Hya. Two orbits are plotted for better visualization.}
\label{fig:vel_emission_ce315}
\end{center}
\end{figure}

\begin{figure}
\begin{center}
\includegraphics[width=0.48\textwidth]{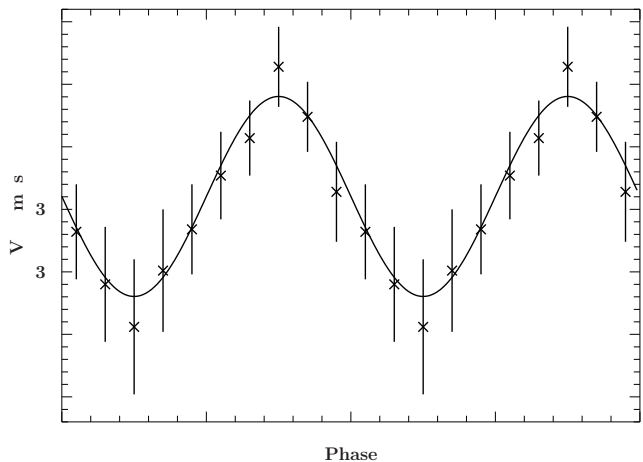}
\caption{Radial velocity curve with measured velocities of the nitrogen absorption line N\,{\sc i} 4143\,\AA\,in GP\,Com. Two orbits are plotted for better visualization.}
\label{fig:absorp_vel}
\end{center}
\end{figure}

\subsection{The orbital ephemeris}

\citet{mar99} and \citet{mor03} have demonstrated that the central-spike in GP Com shows significant radial velocity shifts as a function of the orbital phase, with an amplitude of $\sim 10-15$\,km\,s$^{-1}$, compatible with its likely origin close to the accretor. 

We perform multi-Gaussian fits to the individual line profiles, modelling the line as a combination of two broad Gaussians representing the double-peaked disc emission plus a narrow Gaussian for the spike. As a first step, we fit the velocities of the central-spike to the individual spectra. The derived radial velocity curves are then used to determine the orbital phasing of the spike.

Although the UVES data and X-Shooter data were taken eight years apart the orbital period could not be refined because the X-Shooter data only cover about one orbit of GP\,Com. Therefore, for GP\,Com, we fix the orbital period to 46.57 minutes \citep{mar99}. If the radial velocity curve of the spike traces the white dwarf accretor, we can define orbital phase zero as the phase of superior conjunction of the accretor  (or the red to blue crossing point of the accretor's radial velocity curve). Individual lines give identical zero-points for the phasing, we thus fit to the three strongest helium lines simultaneously, and derive the following ephemeris for GP\,Com taking either the UVES or the X-Shooter data:

\[ \mathrm{HJD}_\mathrm{GP\,Com; UVES} = 2452372.5994(2) + 0.0323386 E \]

\[ \mathrm{HJD}_{\rm GP\,Com; X{\textrm-}Shooter} = 2455245.851(2) + 0.0323386 E \]


\begin{figure*}
\begin{center}
\includegraphics[width=0.85\textwidth]{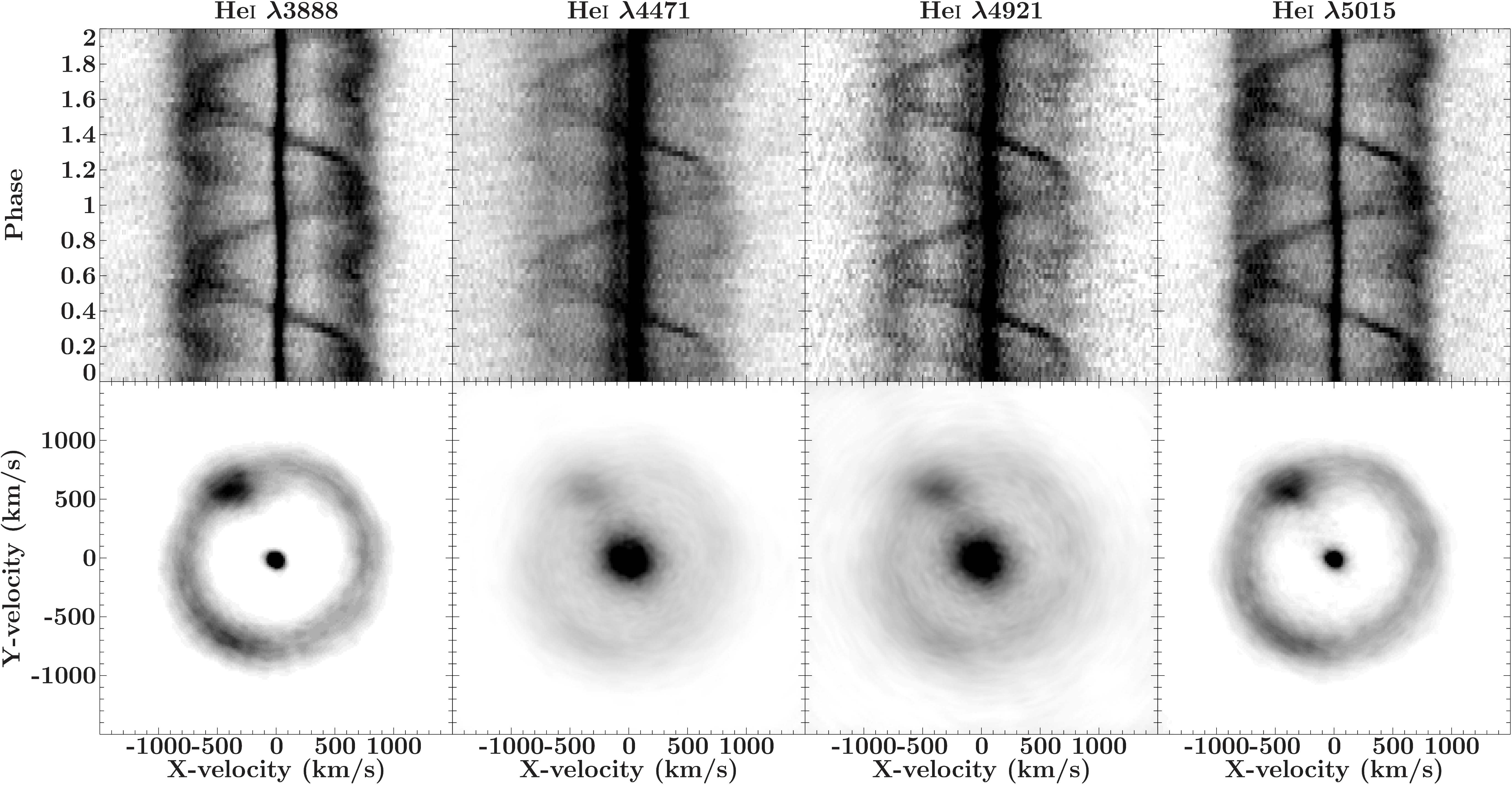}\\
\vspace{0.3cm}
\includegraphics[width=0.85\textwidth]{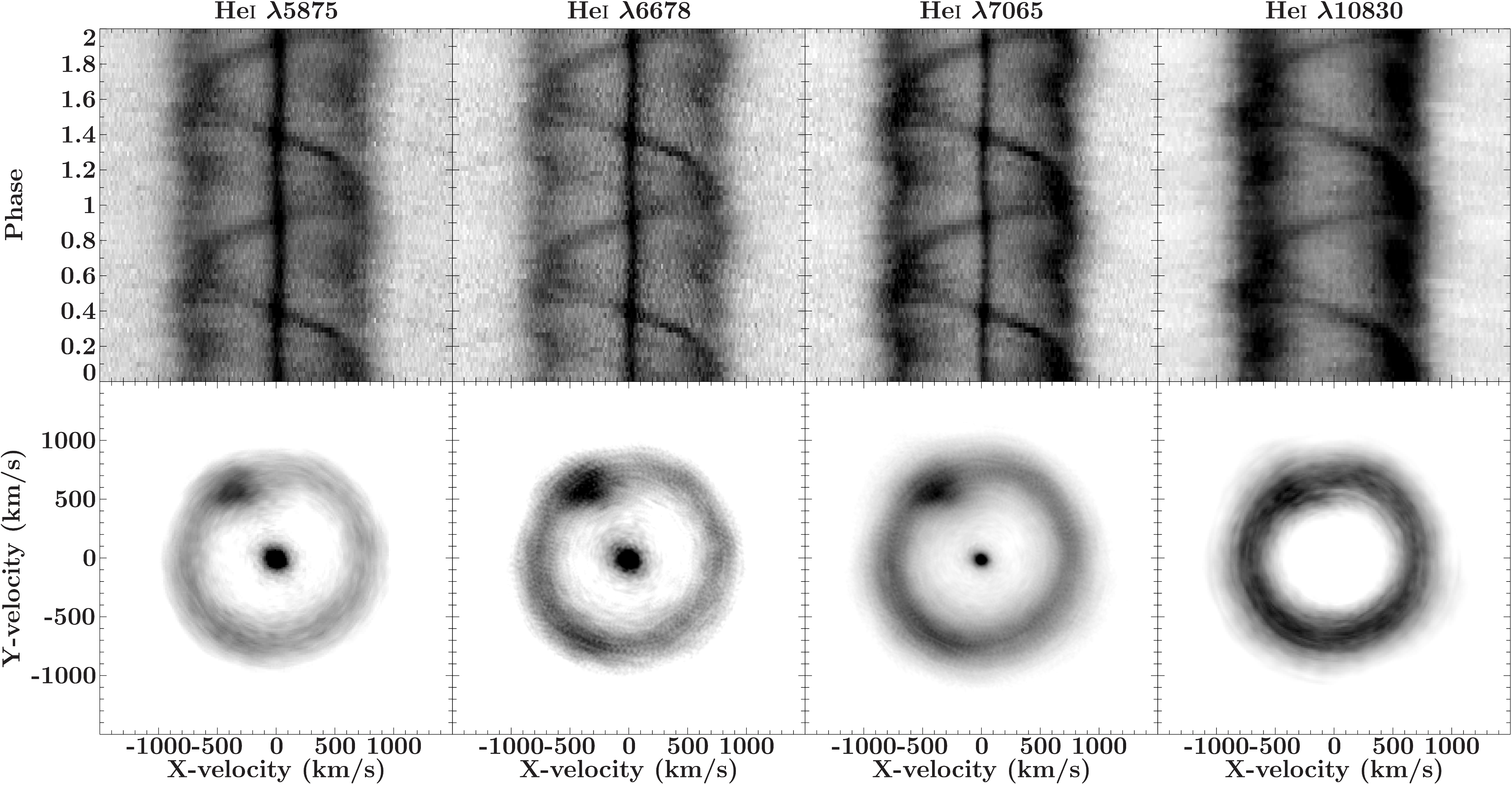}
\caption{Trailed spectra and maximum-entropy Doppler tomograms of selected He\,{\sc i} lines of GP\,Com obtained from X-Shooter data. The disc, the central-spike as well as two bright spots are visible. Note that in the plotting of some lines the displayed intensity of the central-spike was saturated to emphasize both bright spots.}
\label{fig:trail_gpcom}
\end{center}
\end{figure*}

\begin{figure*}
\begin{center}
\includegraphics[width=0.85\textwidth]{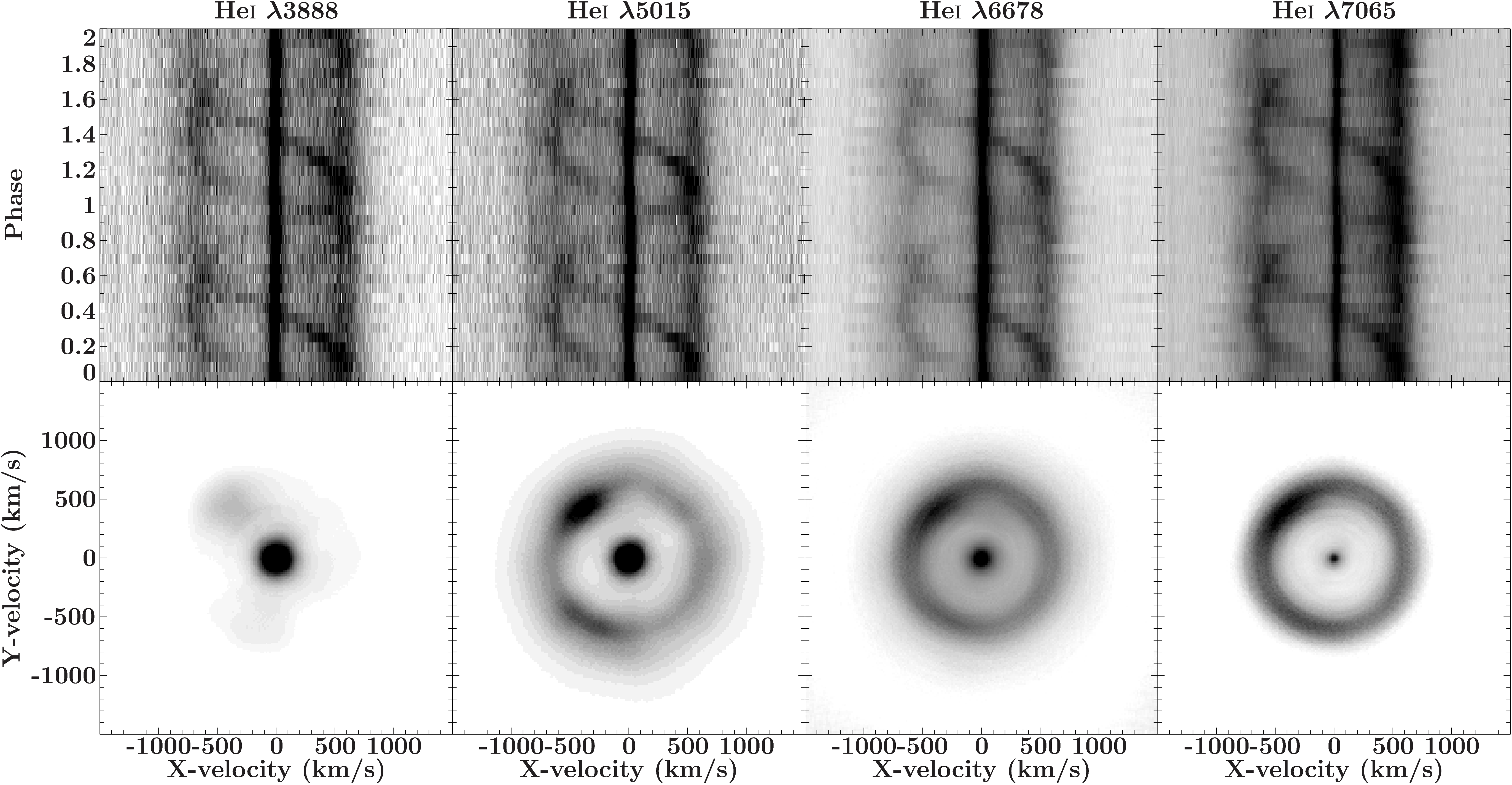}
\caption{Trailed spectra and maximum-entropy Doppler tomograms of selected He\,{\sc i} lines of V396 Hya obtained from the UVES data. The disc, the central-spike as well as both bright spots are visible. Note that in some lines the central-spike was saturated to emphasize both bright spots.}
\label{fig:trail_ce315}
\end{center}
\end{figure*}

In the case of V396\,Hya, the orbital period is not that well constrained, but the short duration of our observations prevents an improvement on the value of 65.1 minutes as derived by \citet{rui01}. The X-Shooter data could not be used because the exposure time of the individual spectra was 15\,min which covers a significant fraction of the orbit. We again use the radial velocity curve of the spike in the three strongest lines to derive the following ephemeris for V396 Hya

\[ \mathrm{HJD}_\mathrm{V396\,Hya; UVES} = 2452372.5263(3) + 0.0452083 E \]

Armed with the orbital ephemerides, we fold all spectra in order to obtain higher signal-to-noise spectra, permitting us to derive more accurate values for the radial velocity curve of the various emission line components. All orbital phases reported in this paper are based on the above ephemerides.

\subsection{The radial velocity of the central-spike and the metal lines}

The final radial velocity curves of the central-spike components are measured by fitting a single Gaussian to the central-spikes in the phase-folded spectra. A pure sinusoid was fitted to the radial velocities to determine the mean and amplitude of the orbital motion of the spike and their formal errors. Table\,\ref{spikevelocities} lists the properties of the spike in all the helium lines for which a reliable fit could be made, for both our targets.

We find that all 12 usable helium lines covered by the UVES spectra are consistent with the same radial velocity amplitude for the central-spike component and thus we calculate a weighted mean from the 12 fitted amplitudes to derive $K_\mathrm{spike} = 11.7 \pm 0.3$\,km\,s$^{-1}$ for GP Com and $K_\mathrm{spike} = 5.8 \pm 0.3$\,km\,s$^{-1}$ for V396 Hya. The fact that all lines share the same phasing and velocity amplitudes provides strong support for placing the origin of the spike at or near the primary white dwarf, thus providing us with an accurate determination of its radial velocity amplitude, $K_1$.  However, the mean velocities of the central-spike show variations from line to line, from $-5$ to $+53$\,km\,s$^{-1}$. The seemingly random systemic velocity shifts reported by \citet{mar99} and \citet{mor03} are thus confirmed in our high-resolution data and extended to a larger number of lines. The shifts strongly correlate between the same lines in both stars.

Not only the central-spike feature shows radial velocity variations. We find that the narrow emission lines of neon and nitrogen, as well as the absorption lines in GP\,Com, trace the motion of the central-spike. Figure\,\ref{fig:trail_emis_gpcom} shows a trailed spectrogram of the center of two strong helium emission lines as well as the two strongest nitrogen and neon emission lines in GP Com. A sinusoidal fit was made to the radial velocities of the metal lines. The radial velocity amplitudes and phases are consistent with those of the central-spike feature in the helium lines. For the co-added neon and nitrogen emission lines, we derive a mean amplitude $K_{\rm emission} = 12.3\pm0.5$\,km\,s$^{-1}$ and a mean velocity $\gamma=11.9\pm0.9$\,km\,s$^{-1}$ for GP\,Com (Fig.\,\ref{fig:vel_emission_gpcom}) and $K_{\rm emission} = 5.6\pm0.8$\,km\,s$^{-1}$ and $\gamma=10.2\pm1.6$\,km\,s$^{-1}$ for V396\,Hya (Fig.\,\ref{fig:vel_emission_ce315}).

The same approach was used for the N\,{\sc i} 4143\,\AA\, absorption line in GP\,Com. We find a radial velocity amplitude of $K_{\rm 4143}=8.5\pm2.1$\,km\,s$^{-1}$ with a mean velocity of $\gamma=35.5\pm2.5$\,km\,s$^{-1}$. The radial velocity amplitude is consistent with the amplitude obtained for the central-spikes (Fig.\,\ref{fig:absorp_vel}). Therefore, we conclude that the metal emission and absorption lines are linked to the accreting white dwarf as well. 


\subsection{System parameters via Doppler tomography}
Doppler tomography \citep{Mar88} of GP\,Com has been used in the past to study the properties of its accretion disc and bright spot emission \citep{mar99,mor03}. Armed with an accurate estimate for the primary radial velocity $K_1$, we can constrain the mass ratios by comparing gas velocities along ballistic trajectories of the accretion stream to the observed bright spot velocities in a Doppler tomogram, as has been done for several of the AM\,CVn stars that show a central-spike \citep{roe05,roe06}.

Fig.\,\ref{fig:trail_gpcom} and \ref{fig:trail_ce315} show maximum-entropy Doppler tomograms of GP Com and V396 Hya. The central-spikes are fixed to the negative Y-velocity axis, the conventional phase of the accreting white dwarf. A strong bright spot shows up near the expected accretion stream/disc impact region, with a faint secondary bright spot at $\sim$120-degree phase-offset, which has also been observed in other AM\,CVn-type systems \citep{roe05,kup13}. We assume that the strong bright spot corresponds to the first impact point of the accretion stream and the disc, while the weaker bright spots may represent accretion stream overflow and re-impact further downstream. The latter effect has been seen in numerical simulations of accretion discs (M. Wood, in preparation; private communication).

\begin{figure*}
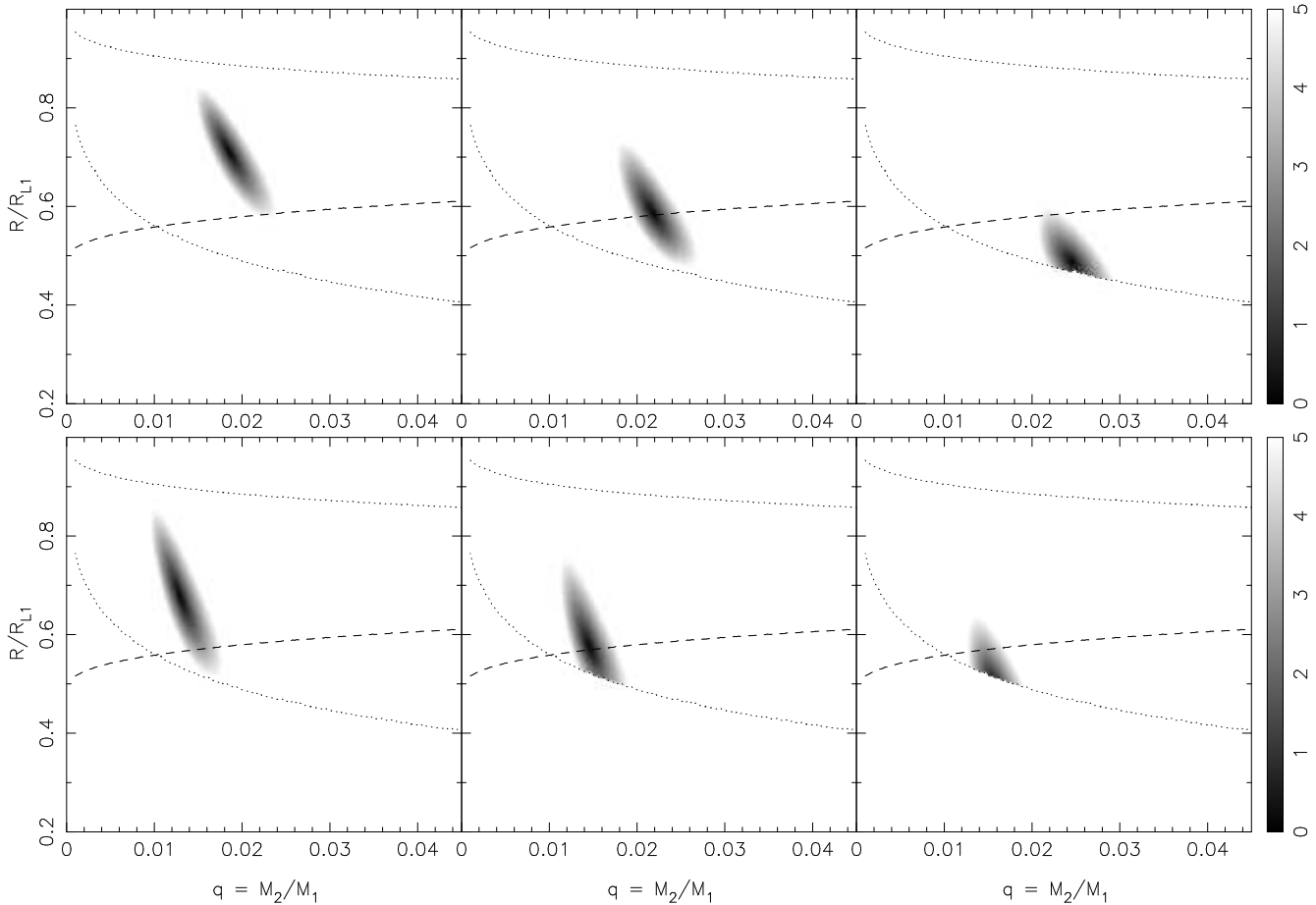

\begin{center}
\includegraphics[angle=-90,width=0.995\textwidth]{constrain_gpcom.ps}
\includegraphics[angle=-90,width=0.995\textwidth]{constrain_v396hya.ps}
\caption{Allowed mass ratios $q$ and effective accretion disc radii $R$ for GP Com (top) and V396 Hya (bottom). The left panels assume ballistic accretion stream velocities in the bright spot, the right panels assume Keplerian disc velocities. Since the bright spot could be a mix of these, we show in the center panels the results for a mix of 80\% ballistic stream velocities and 20\% Keplerian disc velocities. The upper and lower dotted lines indicate the edge of the primary Roche lobe and the circularization radius, respectively, while the dashed line shows the 3:1 resonance radius. The gray-scale indicates the exclusion level in standard deviations.}
\label{systemparams}
\end{center}
\end{figure*}

Figure\,\ref{systemparams} shows the allowed mass ratios and accretion disc radii for GP Com and V396 Hya that we obtain from the phases and amplitudes of the central-spikes and the primary bright spot. We solve the equation of motion for a free-falling stream of matter through the inner Lagrangian point, based on the results of \citet{lub75}, and we see whether the resulting accretion stream and/or accretion disc velocities and phases at the stream-disc impact point match with the measured values. The bright spot in interacting binaries is not always observed to represent the pure ballistic stream velocities at the stream-disc impact point; the gas velocities could in principle lie anywhere between the ballistic stream velocities and the accretion disc velocities in the disc-stream impact region. We therefore consider the two limiting cases of pure ballistic stream and pure Keplerian disc velocities in the bright spot. The mass ratio and accretion disc radii (i.e., where the bright spot occurs) ranges we obtain in this way for GP\,Com are $0.015 < q < 0.022$ and $0.60 < R/R_{L_1} < 0.80$ assuming pure ballistic stream velocity and $0.021 < q < 0.028$ and $0.45 < R/R_{L_1} < 0.58$ assuming pure Keplerian disc velocity. For V396\,Hya we find $0.010 < q < 0.016$ and $0.54 < R/R_{L_1} < 0.84$ assuming pure ballistic stream velocity and $0.013 < q < 0.018$ and $0.50 < R/R_{L_1} < 0.62$ assuming pure Keplerian disc velocity. $R_{L_1}$ corresponds to the distance from the center of the accreting white dwarf to the inner Lagrangian point. 

While these mass ratios and disc radii lie in the range where superhumps might be expected due to the $3:1$ resonance, numerical simulations indicate that the mass ratios are in fact so low that the (eccentric) accretion disc remains stationary in the binary frame \citep{sim98}. This matches the absence of any reports of `superhumps' in GP Com and V396 Hya.

\begin{table}
\begin{center}
\caption{System parameter either assuming pure ballistic stream velocities pure Keplerian disc velocities.}
\begin{tabular}{lll}
\hline
 & pure ballistic stream & pure Keplerian disc \\
 \hline \hline
 \multicolumn{3}{l}{\bf GP\,Com} \\
$q$          &  $0.020$ - $0.022$     &    $0.024$ - $0.028$      \\
$M_\mathrm{1}$ (M$_\odot$)   &  $>0.54$    & $>0.33$   \\
$M_\mathrm{2}$ (M$_{\rm Jupiter}$)   & $12.5$ - $33.8$    & $9.6$ - $42.8$ \\
$R/R_{L_1}$  &   $0.68$ - $0.80$    &     $0.45$ - $0.58$     \\
\smallskip
$i$ ($^\circ$)  & $45$ - $74$   & $33$ - $78$  \\
  {\bf V396\,Hya} &     &         \\
$q$          &   $0.010$ - $0.016$    &   $0.013$ - $0.018$       \\
$M_\mathrm{1}$ (M$_\odot$)  &   $>0.37$   &  $>0.32$  \\
$M_\mathrm{2}$ (M$_{\rm Jupiter}$)   &  $6.1$ - $26.8$   &   $6.1$ - $30.5$ \\
$R/R_{L_1}$  &   $0.54$ - $0.84$    &     $0.50$ - $0.62$     \\
$i$ ($^\circ$)  &  $29$ - $75$  &   $25$ - $79$ \\  
     \hline
\end{tabular}
\label{tab:systempara}
\end{center}
\end{table}  

With the derived mass ratios and the primary velocity amplitude $K_1$ as well as the measured disc velocities we can limit the inclination angles, the component masses and the disc sizes. Above a certain inclination angle the donor will start to eclipse the outer edges of the disc and, at high enough inclination, also the accretor. No eclipses are observed in both systems either in the lines nor in the continuum. 

\begin{figure*}
\begin{center}
        \resizebox{8.5cm}{!}{\includegraphics{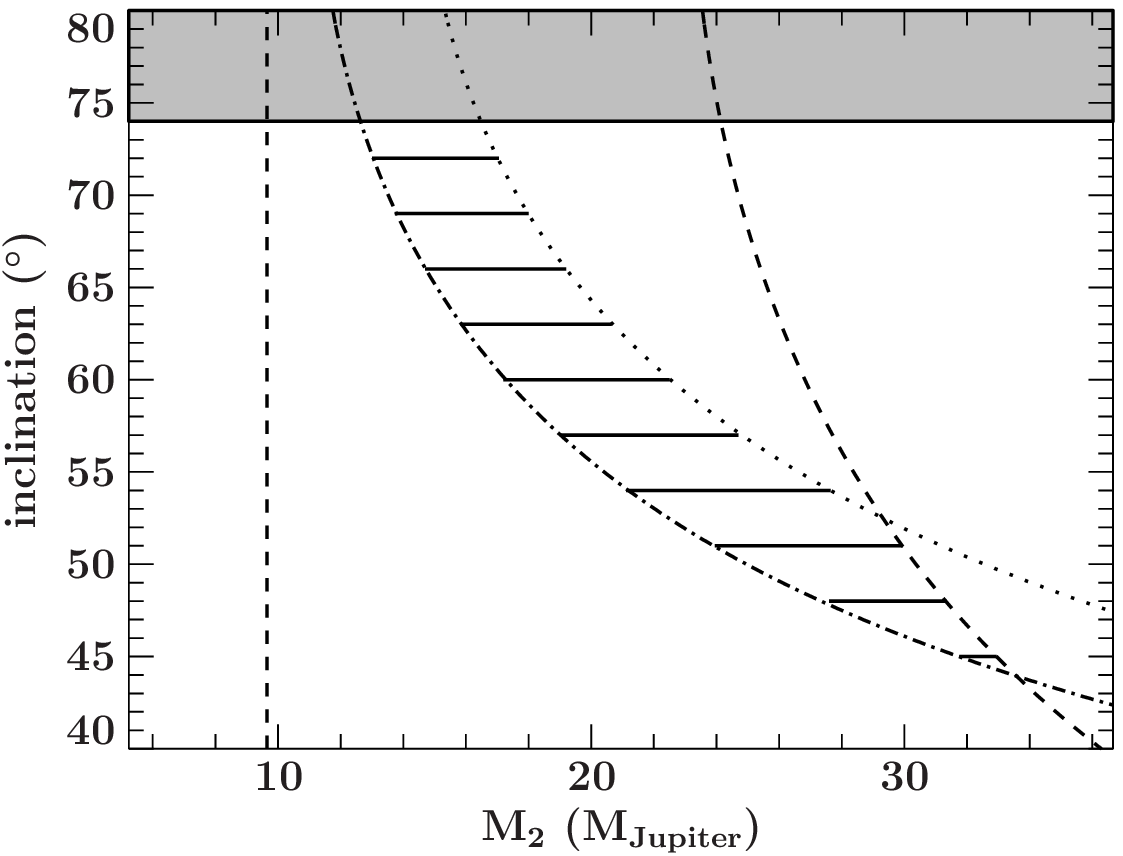}}
         \resizebox{8.5cm}{!}{\includegraphics{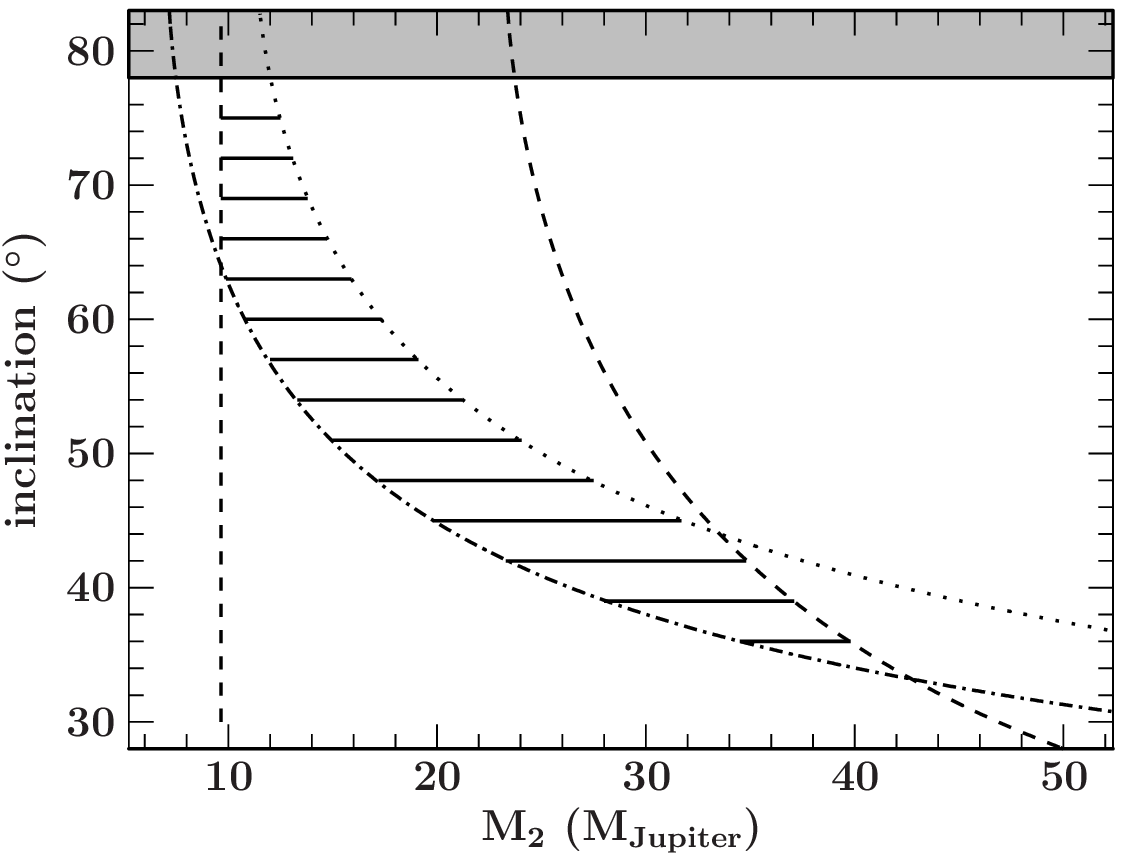}}
         \resizebox{8.5cm}{!}{\includegraphics{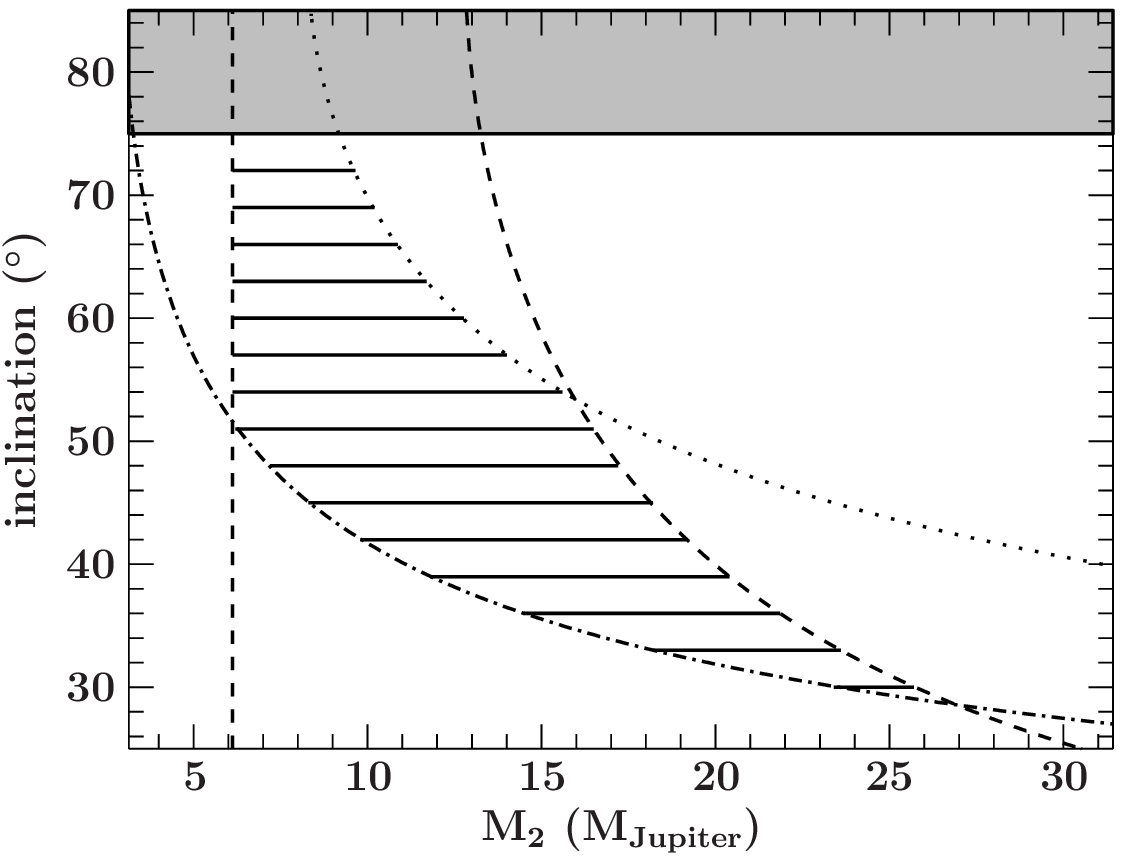}}
          \resizebox{8.5cm}{!}{\includegraphics{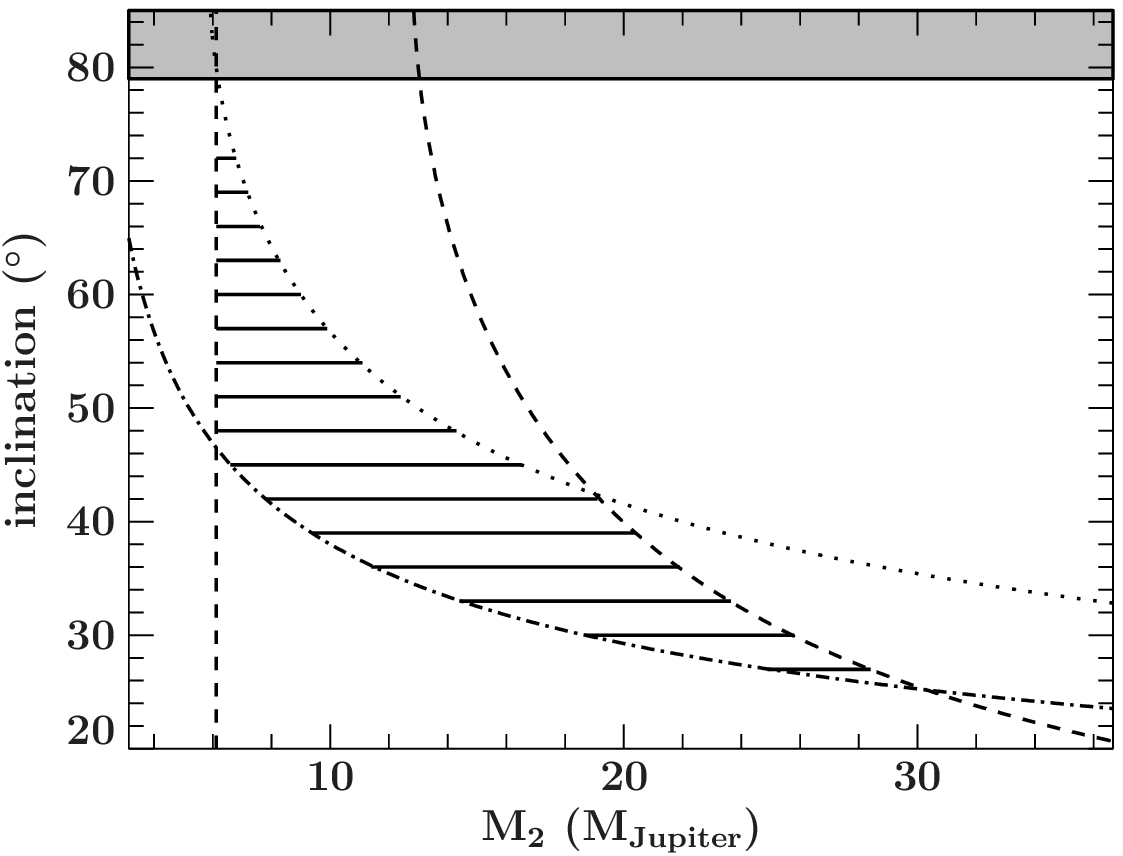}}
         \end{center}
\caption{Calculated inclination angles for different donor masses for GP\,Com (upper panels) and V396\,Hya (lower panels). The left vertical dashed line marks the zero temperature mass for the donor star and the curved dashed line marks the limit when the accretor reaches the Chandrasekhar limit. The grey shaded area marks the region where the system would show eclipses assuming the largest possible disc radius. The dotted line corresponds to the largest disc radius and the dashed-dotted line to the largest mass ratio. The hatched area shows the allowed parameter range. {\bf Upper left:} Assuming pure ballistic stream velocities for GP Com. {\bf Upper right:} Assuming pure Keplerian disc velocities for GP Com. {\bf Lower left:} Assuming pure ballistic stream velocities for V396\,Hya. {\bf Lower right:} Assuming pure Keplerian disc velocities for V396\,Hya. }
\label{fig:mass_incl}
\end{figure*}       

For a given mass ratio and donor mass we calculate the expected primary velocity amplitudes. The ratio of the calculated velocity amplitudes and the measured velocity amplitudes depends only on the inclination angle. Additionally, the measured peak-to-peak (Keplerian) velocities in the double-peaked lines correlate to the velocity in the outer disc, and also depend on the inclination in a similar fashion. \citep{hor86a}. 

Fig.\,\ref{fig:mass_incl} shows the calculated inclination angles for different donor masses for GP\,Com (upper panels) and V396\,Hya (lower panels) assuming either pure ballistic stream or pure Keplerian velocity of the bright spot. As a lower limit to the mass of the donor, the mass for a Roche lobe filling zero-temperature helium object is calculated \citep{del07}. The upper limit for the donor masses are set by the Chandrasekhar mass of the primary. The dotted line in Fig.\,\ref{fig:mass_incl} corresponds to the largest allowed disc radius ($0.80\,R/R_{L_1}$ for pure ballistic stream velocities and $0.58\,R/R_{L_1}$ assuming pure Keplerian disc velocity in GP\,Com). The largest allowed disc radius is an upper limit. The dashed-dotted line in Fig.\,\ref{fig:mass_incl} corresponds to the largest possible mass ratio ($q=0.022$ for pure ballistic stream velocities and $q=0.028$ assuming pure Keplerian disc velocities in GP\,Com). Smaller mass ratios would shift the dashed dotted lines to the right. Hence, the largest mass ratio is a lower limit. Combining these constraints leaves a small region of parameter space that is allowed. 

The upper panels in Fig.\,\ref{fig:mass_incl} show the allowed ranges in inclination for a given donor mass in GP\,Com and the lower panels in Fig.\,\ref{fig:mass_incl} for V396\,Hya. Table\,\ref{tab:systempara} gives an overview of the derived system parameters either assuming pure ballistic stream velocities or pure Keplerian disc velocities.

\section{Discussion}








\begin{figure*}
\begin{center}
\includegraphics[width=14cm]{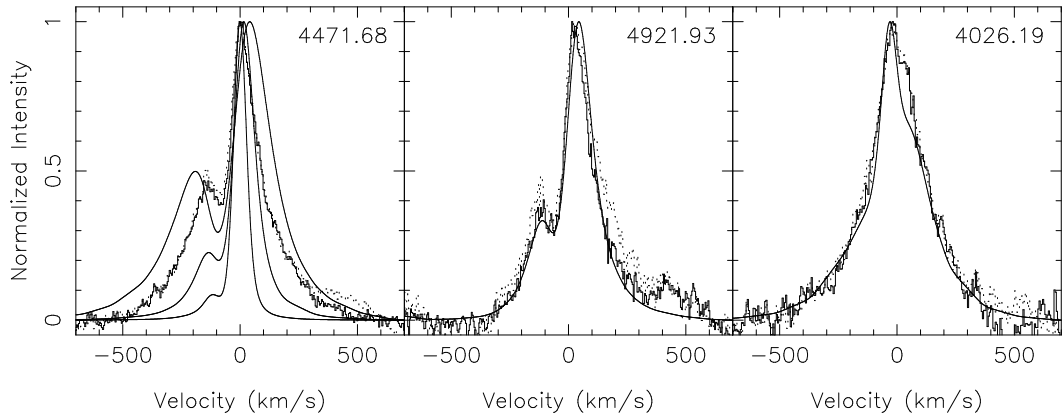}
\caption{Observed spikes from the UVES spectra in GP Com (dotted histogram) and V396 Hya (solid histogram) for the three He\,{\sc i} lines that contain observed forbidden components, together with the modelled Stark-broadened line profiles (solid line) from \citet{bea97}. The left panel shows how the amplitude and displacement of the forbidden component increases with electron density; the models shown are for electron densities of $n_e=1\times 10^{15}$, $3\times 10^{15}$, and $1\times 10^{16}$ cm$^{-3}$. In the middle and right panel we show the model for $n_e = 3\times 10^{15}$ cm$^{-3}$.}
\label{spikefits}
\end{center}
\end{figure*}

\begin{figure*}
\begin{center}
\includegraphics[width=10cm]{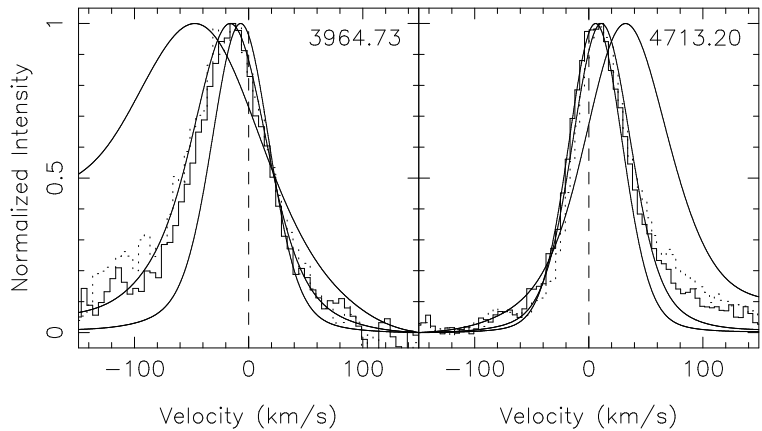}
\caption{Observed spikes from the UVES spectra  in GP Com (dotted histogram) and V396 Hya (solid histogram) for two more He\,{\sc i} lines, together with the modelled Stark-broadened line profiles from \citet{bea97}. The lines were chosen for their relatively large observed shifts. The models shown are for electron densities of $n_e=1\times 10^{15}$, $3\times 10^{15}$, and $1\times 10^{16}$ cm$^{-3}$; the broadening and shift of the lines increases with electron density. All line types are the same as in Fig.\ \ref{spikefits}.}
\label{spikefits2}
\end{center}
\end{figure*}

\subsection{Stark broadening and the behaviour of the central emission spikes}

The forbidden components of neutral helium that are observed in some of the central-spikes (Fig.\ \ref{spikefits}) have led to the suggestion that the central-spike profiles may be affected by Stark broadening \citep{mor03}, as modelled by \citet{bea97} and \citet{bea98} for (single) DB white dwarfs and extreme helium (EHe) stars. Unfortunately those models appeared to predict the wrong splittings between the allowed and forbidden emission features in GP Com, such that questions remained as to whether the Stark broadening explanation was correct \citep{mor03}.

However, as shown in Fig.\,1 of \citet{gri68}, the splitting of allowed and forbidden transitions of neutral helium depends on the electron density $n_e$. The splittings of the absorption lines in those models are therefore relevant for those stellar atmospheres only. The emission lines in GP Com and V396 Hya may be caused by a single-temperature, optically thin layer of helium (see analysis by \citealp{mar91}), such that the splittings between the allowed and forbidden transitions have a different value corresponding to the local electron density in the layer.

In Fig.\,\ref{spikefits} we show the observed profiles of the central-spikes in the He{\sc i} 4471\,\AA\, and He{\sc i} 4921\,\AA\, lines in detail. Both these lines show a forbidden component blueward of the allowed transition, the strength and distance of which (relative to the allowed component) can be correctly reproduced by Stark-broadened line profile models for helium plasma with an electron density close to $n_e\simeq 5\times 10^{15}$ cm$^{-3}$. The models shown in Fig.\ \ref{spikefits} are from \citet{bea97}, as obtained in tabular form from the stellar spectral synthesis program \textsc{spectrum}. As an independent check, Fig.\,1 in \citet{gri68} points to a similar electron density.

In addition to the forbidden neutral helium lines near He{\sc i} 4471\,\AA\,and He{\sc i} 4921\,\AA, there is a blended forbidden component predicted blueward of He{\sc i} 4026\,\AA, which is confirmed by our data and can be seen in the asymmetric line profile of He{\sc i} 4026\,\AA\,(Fig.\,\ref{spikefits}). For the other lines, assuming similar electron densities in the plasma, the models do not predict observable satellite spikes, but they do predict shifts of the emission lines by various amounts relative to their rest wavelengths. Both blueshifts and redshifts are expected, up to a few tens of km\,s$^{-1}$. Fig.\,\ref{spikefits2} shows two helium lines where different shifts for different electron densities are expected. In Table \ref{spikevelocities} we see that the observed spikes are predominantly redshifted; however, the observed redshifts of the central-spike can be matched reasonably well with the expected blue- and redshifts due to Stark broadening if we assume an additional, global redshift of about 15\,km\,s$^{-1}$ which is a combination of the systemic velocity and the gravitational redshift. The spikes shown in Fig.\,\ref{spikefits} were blueshifted by 15\,km\,s$^{-1}$. Unlike the neutral helium lines, the hydrogenic He{\sc ii} 4686\,\AA\, line should only be broadened and \emph{not} shifted due to the Stark effect: its measured redshift should therefore match the global redshift found in the He{\sc i} lines. With a measured redshift of $17.4\pm0.3$ for GP\,Com and $16.1\pm0.5$ km\,s$^{-1}$ for V396\,Hya (Table \ref{spikevelocities}), this is indeed the case. 



The emission spikes in the AM\,CVn stars may provide a unique view of the effects of Stark broadening on the appearance of forbidden lines of helium. Whereas such lines in DB white dwarf or EHe star atmospheres are always a convolution of an entire stellar atmosphere, the lines in AM\,CVn stars appear to be produced in an optically thin slab of gas with a single density, or very narrow range of densities. Model predictions for Stark-broadened and forbidden helium lines further to the UV could be tested by our X-Shooter spectroscopy. Our fitting of the observed Stark profiles with calculated profiles is currently limited by the resolution of the grid of models available to us.










\begin{figure}
\begin{center}
\includegraphics[width=0.48\textwidth]{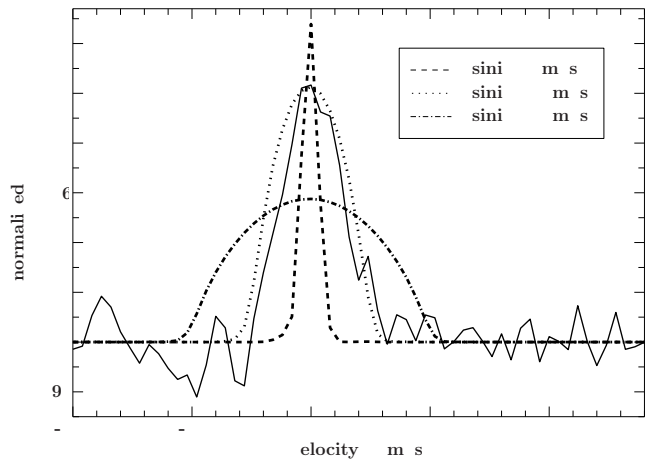}
\caption{The average line profile of the narrow neon emission lines of GP\,Com from UVES spectra with computed line profiles for different rotational velocities.}
\label{rot}
\end{center}
\end{figure}

\subsection{Metal lines}

\citet{kup13} discovered strong absorption lines of magnesium and silicon in the three AM\,CVn systems known to have orbital periods between $50$-$60$\,min. However, they were not able to trace the origin of these lines because of the low spectral-resolution data. We show that the nitrogen absorption lines in GP\,Com follow the motion of the central-spike. Therefore, the origin of the absorption lines are most likely photospheric absorption lines from the accretor which is accreted material from the disc. That means that the observed abundances most likely represent the composition of the disc and therefore the donor star.

We have identified a large number of nitrogen and neon lines as well as the O\,{\sc i} multiplet at 7771/74/75\,\AA\,in GP\,Com. However no evidence for iron and silicon can be found in either system. The strength of iron and silicon lines could be used to determine the initial metalicity since their abundance is not supposed to be affected by nuclear synthesis processes during binary evolution. \citet{mar91} concluded that the non-detection of heavier elements such as iron and silicon shows that GP\,Com is a low metallicity object. The strong enhancement of nitrogen is a clear sign of CNO-processed material \citep{nel10}. This seems also to apply to V396\,Hya. More intriguing is the overabundance of neon which was also found in X-ray observations of GP\,Com \citep{str04} because the only way to enhance neon is during helium burning, which decreases the nitrogen abundance and increases the carbon and oxygen abundance.

If the observed abundances represent the composition of the donor, the overabundance of neon can only be explained through a short phase of helium burning which stopped before nitrogen is depleted and carbon and oxygen is enhanced significantly \citep{arn99}. We do not see an enhanced abundance of oxygen and carbon, therefore a formerly highly-evolved helium star donor is excluded in both systems. 

As an alternative scenario the observed abundances in the donor star could be affected by crystallization processes in the core. \citet{yun02} found that a crystallized and fractionated core could lead to strong ${}^{22}$Ne enhancement. For white dwarfs these processes would take several Gyr which is similar to the expected age of GP\,Com and V396\,Hya. This would explain the large overabundance of neon, although also here some helium burning occurred in the donor star.

\begin{figure}
\begin{center}
\includegraphics[width=0.48\textwidth]{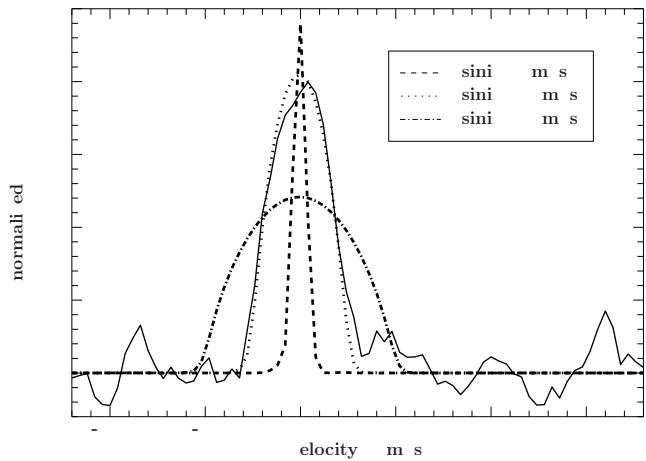}
\caption{The average line profile of the narrow neon emission lines of V396\,Hya from UVES spectra with computed line profiles for different rotational velocities.}
\label{rot_ce315}
\end{center}
\end{figure}

\subsection{Rotational velocity of the accretor}
\citet{mar04} showed that the synchronization torque between the accretor and the orbit can feed back angular momentum into the orbit which possibly spins down the accretor, stabilizes the orbit and prevents the merger of the system. This will have a positive influence on the formation rate of AM\,CVn systems if the synchronization timescale is short enough. To put further constraints on the synchronization timescales the rotational period of the accretor has to be compared to the orbital period of the binary. Despite the lower SNR, the narrow neon lines are not affected by Stark broadening and therefore these lines can be used to estimate the rotational velocity of the accretor.

The narrow neon and nitrogen emission lines are compared to synthetic profiles from a single slab LTE model with uniform temperature and density which was also used by \citet{mar91} for their abundance analysis on GP\,Com. The computed line profile includes only thermal broadening. Additional broadening effects like microturbulence and Stark broadening are neglected. A synthetic model was computed using the parameters derived for GP Com \citep{mar91}. Additional rotational broadening following \citet{gra08} was applied to the synthetic models. 

Fig.\,\ref{rot} and \ref{rot_ce315} shows that a broadening of $v_{\rm rot}=25$\,km\,s$^{-1}$ is needed to find good agreement between the models and the observed emission lines in GP\,Com and V396\,Hya. This value is only a lower limit and depends on the inclination angle. However, even for a maximum accretor mass, we can set a limiting rotational velocity of $v_{\rm rot}<46$\,km\,s$^{-1}$ for GP\,Com and $v_{\rm rot}<59$\,km\,s$^{-1}$ for V396\,Hya. Therefore, we conclude that the accretor does not rotate fast in both systems and must have spun down significantly during the evolution of GP Com and V396\,Hya as AM\,CVn-type systems.      

If the accreting white dwarf is tidally locked to the orbit, the rotational velocity of the accretor can be calculated using the following equation:
\begin{equation}
v_{\rm rot}=\frac{2 \pi R_{\rm WD}}{P_{\rm orb}}
\end{equation}
where $P_{\rm orb}$ corresponds to the orbital period of the system and $R_{\rm WD}$ to the radius of the accreting white dwarf. A radius of the white dwarf accretor in both systems was estimated to be $R_{\rm WD}=0.015$\,R$_\odot$ using the zero-temperature mass-radius relation of Eggleton (quoted in \citealt{ver88}) for an $0.5$\,M$_\odot$ white dwarf. This leads to a synchronized equatorial rotational velocity of $v_{\rm rot; GP Com}=23.4$\,km\,s$^{-1}$ and $v_{\rm rot; V396 Hya}=16.8$\,km\,s$^{-1}$ for the accretor in GP\,Com and V396\,Hya respectively. This is only an upper limit as the observed rotational velocity depends on the inclination angle of the system. 

Although we can exclude fast rotational velocities, we cannot draw a firm conclusion whether the accretor is tidally locked. 

\section{Conclusions and summary}
The average spectra of GP\,Com and V396\,Hya reveal strong double-peaked helium emission lines typical for long period AM\,CVn-type systems. All the lines in the optical regime show, on top of the helium disc emission lines, a strong central-spike feature. Interestingly, none of the helium lines in the near-infrared of GP\,Com and V396\,Hya show this sharp central-spike feature. 

Besides the strong helium lines a large number of narrow nitrogen and neon emission lines are detected in both systems. Additionally broad nitrogen absorption lines are also detected in GP\,Com. We show that the neon and nitrogen lines follow the motion of the central-spike and therefore have an origin on or close to the accretor with the absorption lines originating most likely in the photosphere of the accreting white dwarf. No evidence for iron and silicon was found which indicates that GP\,Com and V396\,Hya are low metallicity objects with an overabundance of neon and nitrogen. The neon and nitrogen lines were found to be connected to the accreting white dwarf and represent the abundance pattern of the accreted material of the donor star. An enhancement of nitrogen can be explained with CNO burning whereas neon is only produced during helium burning where nitrogen is burned into carbon and oxygen. Therefore, we find no satisfying solution to explain both high neon and high nitrogen and can only exclude a highly evolved helium star donor in both systems.

We find a large variation of the mean velocities of the central-spike features ranging from $-5$ to $+53$\,km\,s$^{-1}$ and detect forbidden components of several helium lines. Stark broadened models predict both the appearance of forbidden helium lines and the displacement for the other helium lines. Therefore, the helium lines are compared to calculated line profiles and we find that the central-spike features can be correctly reproduced by Stark-broadened line profile models for helium plasma with an electron density close to $n_e\simeq 5\times 10^{15}$ cm$^{-3}$.

The small semi-amplitude of the central-spike, the consistency of phase and amplitude with the absorption components in GP\,Com as well as the measured Stark effect shows that the central-spike originates on the accreting white dwarf and can therefore be used to trace the motion of the accretor.\\ 

Doppler tomograms reveal a strong accretion disc bright spot with a faint secondary spot at an $\sim$120-degree offset to the first bright spot. From the central-spikes and primary bright spot we limit the inclination angles and the component masses. For GP\,Com, we find component masses of $M_\mathrm{1,GP\,Com}>0.33$\,M$_\odot$ and $M_\mathrm{2,GP\,Com} = 9.6$ - $42.8$\,M$_{\rm Jupiter} $ seen under an inclination angle of $33^\circ<i<78^\circ$. For V396\,Hya we find a possible mass for the accretor of $M_\mathrm{1,V396\,Hya}>0.32$\,M$_\odot$ with a companion of $M_\mathrm{2,V396\,Hya} = 6.1$ - $30.5$\,M$_{\rm Jupiter}$ and the system is seen under an inclination angle of $25^\circ<i<79^\circ$.



By comparing the line profile of the metal lines with model spectra, we show that the lines can be reproduced with a projected rotational broadening of $v_{\rm rot}\sim25$\,km\,s$^{-1}$ and therefore conclude that rotational velocity of the accreting white dwarf has to be $v_{\rm rot}<46$\,km\,s$^{-1}$ for GP\,Com and $v_{\rm rot}<59$\,km\,s$^{-1}$ for V396\,Hya which excludes fast rotating accretors in both GP\,Com and V396\,Hya.   

\section*{Acknowledgments}
TK acknowledges support by the Netherlands Research School for Astronomy (NOVA). TRM and DS acknowledge the support from the Science and Technology Facilities Council (STFC),  ST/L00733, during the course of this work. We thank Simon Jeffery for useful discusison on the spectral features.

\bibliography{refs}{}
\bibliographystyle{mn2e}

\appendix
\section{}

\begin{table}
\begin{center}
\caption{Equivalent widths of the nitrogen spectral lines}
\begin{tabular}{lll}
\hline
& GP Com & V396 Hya  \\
line & EW (\AA) &  EW (\AA)   \\
\hline\hline
N\,{\sc ii}\,3995 & --0.06$\pm$0.01 & --0.08$\pm$0.01  \\
N\,{\sc i}\,4099 & 0.53$\pm$0.02 & --0.04$\pm$0.01  \\ 
N\,{\sc i}\,4109 & 0.40$\pm$0.02 & --0.08$\pm$0.01  \\
N\,{\sc i}\,4137 & 0.10$\pm$0.01 &  -$^f$ \\
N\,{\sc i}\,4143 & 0.14$\pm$0.02 & --0.71$\pm$0.02$^d$ \\
N\,{\sc i}\,4151 & 0.32$\pm$0.02 &   \\
N\,{\sc i}\,4214/15 & 0.20$\pm$0.01 &  - \\
N\,{\sc i}\,4223/24 & 0.40$\pm$0.02 & -  \\
N\,{\sc i}\,4230 & 0.11$\pm$0.01 & -  \\
N\,{\sc i}\,4253 & 0.14$\pm$0.02 & -  \\
N\,{\sc i}\,4336 & 0.06$\pm$0.01 & -  \\
N\,{\sc i}\,4342 & 0.11$\pm$0.01 & -  \\
N\,{\sc i}\,4342 & 0.11$\pm$0.02 & -  \\
N\,{\sc i}\,4358 & 0.21$\pm$0.02 & -  \\
N\,{\sc ii}\,4447 & --0.03$\pm$0.01 & --0.15$\pm$0.02  \\
N\,{\sc ii}\,4630 & --0.06$\pm$0.01 &  --0.08$\pm$0.01 \\
N\,{\sc i}\,5281 & 0.16$\pm$0.02 & -  \\
N\,{\sc i}\,5292 & 0.07$\pm$0.01 & -  \\
N\,{\sc i}\,5310 & 0.06$\pm$0.01 & -  \\
N\,{\sc i}\,5328 & 0.12$\pm$0.02 & -  \\
N\,{\sc i}\,5356 & 0.07$\pm$0.01 & -  \\
N\,{\sc ii}\,5666 & --0.05$\pm$0.01 & $^a$  \\
N\,{\sc ii}\,5676 & --0.03$\pm$0.01 & -  \\
N\,{\sc ii}\,5679 & --0.10$\pm$0.01 & --0.05$\pm$0.01  \\
N\,{\sc ii}\,5686 & --0.02$\pm$0.01 & -  \\
N\,{\sc i}\,5999 & 0.15$\pm$0.02 & --0.06$\pm$0.01  \\
N\,{\sc i}\,6008 & 0.21$\pm$0.02 & --0.07$\pm$0.01  \\
N\,{\sc ii}\,6482 & $^b$ &  --0.71$\pm$0.03$^b$  \\
N\,{\sc i}\,6644 & $^a$  & --0.21$\pm$0.02$^e$  \\
N\,{\sc i}\,6646 & $^a$  &   \\
N\,{\sc i}\,7423 & --0.15$\pm$0.02  & --0.49$\pm$0.02  \\
N\,{\sc i}\,7442 & --0.23$\pm$0.02  & --0.79$\pm$0.02  \\
N\,{\sc i}\,7468 & --0.27$\pm$0.02 & --0.94$\pm$0.02  \\
N\,{\sc i}\,7899 & --0.06$\pm$0.01  & --0.42$\pm$0.02  \\
N\,{\sc i}\,8184 & --0.50$\pm$0.02  & --0.92$\pm$0.02  \\
N\,{\sc i}\,8188 & --0.52$\pm$0.02  & --0.85$\pm$0.02  \\
N\,{\sc i}\,8200/01 &   --0.31$\pm$0.02     & --0.81$\pm$0.02  \\
N\,{\sc i}\,8210 &   --0.21$\pm$0.02     & --0.58$\pm$0.02  \\
N\,{\sc i}\,8216 & --0.72$\pm$0.02  & --1.51$\pm$0.03  \\
N\,{\sc i}\,8223 & --0.45$\pm$0.02  & --0.84$\pm$0.02  \\
N\,{\sc i}\,8242 & --0.29$\pm$0.02  & --0.76$\pm$0.02  \\
N\,{\sc i}\,8567 & --0.08$\pm$0.01  & --0.52$\pm$0.02   \\
N\,{\sc i}\,8594 & --0.13$\pm$0.01  & --0.72$\pm$0.02  \\
N\,{\sc i}\,8629 & $^c$  & $^c$  \\
N\,{\sc i}\,8655 & $^c$  & $^c$  \\
N\,{\sc i}\,8680/83/86 & --2.93$\pm$0.03  & --4.76$\pm$0.04  \\
N\,{\sc i}\,8703 & --0.32$\pm$0.03  & --1.09$\pm$0.02  \\
N\,{\sc i}\,8711 & --0.38$\pm$0.03  & --1.21$\pm$0.03  \\
N\,{\sc i}\,8718 & --0.37$\pm$0.03  & --1.00$\pm$0.03  \\
N\,{\sc i}\,8728 &   -  & -0.52$\pm$0.02  \\
N\,{\sc i}\,8747 &   -  & -0.21$\pm$0.02  \\
N\,{\sc i}\,9028 & --0.28$\pm$0.02  & --0.69$\pm$0.02  \\
N\,{\sc i}\,9045 & --0.48$\pm$0.03  & --0.99$\pm$0.02  \\
N\,{\sc i}\,9060 & --0.34$\pm$0.03  & --0.88$\pm$0.02  \\
N\,{\sc i}\,9187 & --0.58$\pm$0.03  & --0.86$\pm$0.02  \\
N\,{\sc i}\,12469 & -  & $^a$  \\
\hline
\end{tabular}
\label{tab:equi_1}
\begin{flushleft}
$^a$ Line present but insufficient SNR to measure\\
$^b$ Blended with N\,{\sc i}\,6481/82/83/84\,\AA\\
$^c$ Line present but contaminated with atmosphere\\
$^d$ Blended with N\,{\sc i}\,4151\,\AA\\
$^e$ Blended with N\,{\sc i}\,6646\,\AA\\
$^f$ Lines marked with a - are below the detection limit of $\sim$0.01\AA
\end{flushleft}
\end{center}
\end{table}

\begin{table}
\begin{center}
\caption{Equivalent widths of spectral lines other than nitrogen}
\begin{tabular}{lll}
\hline
& GP Com & V396 Hya  \\
line & EW [\AA] &  EW [\AA]   \\
\hline\hline
He\,{\sc ii}\,3203 & --0.08$\pm$0.01 & --0.34$\pm$0.03 \\
unidentified 6460 & 0.09$\pm$0.01  &  -$^b$  \\
unidentified 6470 & 0.25$\pm$0.02  &  - \\
Ne\,{\sc i}\,5656 & --0.03$\pm$0.01 & -  \\
Ne\,{\sc i}\,5852 & - &  $^a$ \\
Ne\,{\sc i}\,6074 & $^a$ & --0.05$\pm$0.01  \\
Ne\,{\sc i}\,6096 & --0.02$\pm$0.01 & --0.09$\pm$0.01  \\
Ne\,{\sc i}\,6143 & --0.04$\pm$0.01 & --0.09$\pm$0.01  \\
Ne\,{\sc i}\,6163 & $^a$ & $^a$  \\
Ne\,{\sc i}\,6266 & --0.03$\pm$0.01 & --0.09$\pm$0.01  \\
Ne\,{\sc i}\,6334 & --0.03$\pm$0.01 & --0.09$\pm$0.01  \\
Ne\,{\sc i}\,6402 & --0.11$\pm$0.01 &  --0.28$\pm$0.02 \\
Ne\,{\sc i}\,6506 & --0.06$\pm$0.01 &  --0.14$\pm$0.02 \\
Ne\,{\sc i}\,6532 & --0.02$\pm$0.01 &  --0.08$\pm$0.02 \\
He\,{\sc ii}\,6560 & --0.22$\pm$0.02 & --0.26$\pm$0.02  \\
Ne\,{\sc i}\,6598 & - &  --0.05$\pm$0.01 \\
Ne\,{\sc i}\,7032 & --0.10$\pm$0.01 & --0.09$\pm$0.01  \\
O\,{\sc i}\,7771/74/75 & --0.04$\pm$0.02 & $^a$  \\
O\,{\sc i}\,8446 & --0.03$\pm$0.02 & -  \\
\hline
\end{tabular}
\label{tab:equi_2}
\begin{flushleft}
$^a$ Line present but insufficient SNR to measure\\
$^b$ Lines marked with a - are below the detection limit of $\sim$0.01\AA
\end{flushleft}
\end{center}
\end{table}

\label{lastpage}

\end{document}